\documentclass[fleqn,10pt]{wlscirep}
\usepackage[utf8]{inputenc}
\usepackage{graphicx, caption, subcaption,array, multirow,soul}
\newcommand{\etal}{\emph{et al. }}
\newcommand{\FTWA}{\mathcal{F}^{\mathbf{TW}}_{A \cup B:A}}
\newcommand{\FTWB}{\mathcal{F}^{\mathbf{TW}}_{A \cup B:B}}
\newcommand{\FGA}{\mathcal{F}^{\mathbf{KOL}}_A}
\newcommand{\FGB}{\mathcal{F}^{\mathbf{KOL}}_B}
\usepackage{todonotes}

\usepackage{mathtools}
\usepackage{comment}
\newcommand{\centered}[1]{\begin{tabular}{l} #1 \end{tabular}}

 \title{Best of Both Worlds: Enforcing Detailed Balance in Machine Learning Models of Transition Rates}

 \author[1,*]{Anjana Anu Talapatra}
\affil[1]{Materials Science and Technology Division, Los Alamos National Laboratory, Los Alamos, New Mexico-87545, USA}
\affil[2]{Present address: GE Research, Schenectady, NY 12309, USA}
\affil[3]{Computer, Computational and Statistical Sciences Division, Los Alamos National Laboratory, Los Alamos, New Mexico-87545, USA}
\affil[4]{X-Computational Physics Division,Los Alamos National Laboratory, Los Alamos, New Mexico-87545, USA}
\affil[5]{Theoretical Division,Los Alamos National Laboratory, Los Alamos, New Mexico-87545, USA}
\affil[*]{Corresponding author: Anjana Anu Talapatra, atalapatra@lanl.gov}
\author[1]{Anup Pandey}
\author[4]{Matthew S. Wilson}
\author[3]{Ying Wai Li}
\author[1,2]{Ghanshyam Pilania}
\author[1]{Blas Pedro Uberuaga}
\author[5]{Danny Perez}


\begin{abstract}
  The slow microstructural evolution of materials often plays a key role in determining material properties.
  When the unit steps of the evolution process are slow, direct simulation approaches such as molecular dynamics become prohibitive and Kinetic Monte-Carlo (kMC) algorithms, where the state-to-state evolution of the system is represented in terms of a continuous-time Markov chain, are instead frequently relied upon to efficiently predict long-time evolution.
  The accuracy of kMC simulations however relies on the complete and accurate knowledge of reaction pathways and corresponding kinetics. This requirement becomes extremely stringent in complex systems such as concentrated alloys where the astronomical number of local atomic configurations makes the {\em a priori} tabulation of all possible transitions impractical.
   Machine learning models of transition kinetics have been used to mitigate this problem by enabling the efficient on-the-fly prediction of kinetic parameters. In this study, we show how 
   physics-informed ML architectures can exactly enforce the detailed balance condition, by construction. Using the diffusion of a vacancy in a concentrated alloy as an example, we show that 
   such ML architectures also exhibit superior performance in terms of prediction accuracy, demonstrating that the imposition of physical constraints can facilitate the accurate learning of 
   barriers at no increase in computational cost.
\end{abstract}

\begin{document}
\maketitle

\thispagestyle{empty}

\section*{Introduction}
Many key processes controlling defect evolution in materials are associated with the crossing of large energy barriers, and hence occur on very long timescales. This often makes the simulation of microstructure evolution, which is critical to the prediction of many material properties, prohibitively expensive using direct methods such as molecular dynamics ~\cite{uberuaga2020computational}. 
A common modeling approach that dramatically reduces the computational cost of long timescale simulations is the so-called kinetic Monte Carlo (kMC) approach~\cite{voter2007introduction}, 
where the evolution of the material is approximated as a continuous time Markov chain expressed on a discrete state space that corresponds to the different long-lived conformations of the system. In addition to the set of possible states, a kMC model also requires the specification of the state-to-state transition rates along all possible reaction pathways. This representation makes it possible to numerically sample long state-to-state trajectories extremely efficiently, thereby enabling long-timescale simulations. The fidelity of kMC simulations with respect to direct molecular dynamics however relies on the completeness of the set of states and transitions included in the model, on the numerical accuracy of the estimated transition rates between states, and on the state-to-state transitions being sufficiently rare~\cite{di2016jump,aristoff2023arbitrarily}. 

In traditional kMC simulations, transition rates are typically expressed in terms of Arrhenius expressions where the prefactors and barriers corresponding to different transition pathways are tabulated {\em a priori}~\cite{voter2007introduction}. In more modern incarnations, \cite{henkelman2001long,beland2011kinetic} transition catalogues are constructed on-the-fly using direct saddle search methods. These approaches are especially powerful for "simple" materials (e.g., elemental solids); however, the combinatorial explosion in the number of required energy barriers for chemically (i.e., concentrated alloys) or topologically (i.e., glassy) complex materials can significantly increase their computational cost. In the following, we concentrate on the problem of simulating the diffusion of defects in chemically complex systems, where the number of topologically-distinct transition pathways is  limited, but the number of possible "decorations" of the initial and final states by different chemical environments is astronomical. As each event requires its own transition rate, {\em a priori} tabulation  is often not feasible, while on-the-fly computation using direct saddle search methods incurs a significant computational cost. In order to address this limitation, machine learning (ML) models have been introduced as valuable alternatives to direct computation, which have the potential to restore high simulation rates and long simulation timescales, while preserving high fidelity.
A number of ML models have been developed to model defect migration energetics in alloys, typically requiring on the order of $10^3 -10^4$ explicit barrier calculations for training \cite{manzoor2021machine, fan2022predicting, xu2022revealing, huang2023high}, which in concentrated alloys can be a very small fraction of the total chemical space, leading to potentially considerable computational speedups.

For example, Pascuet~\etal~\cite{pascuet2011stability} were one of the first to predict the migration barriers of Cu vacancy clusters $V_n$ for 1$\leq$n$\leq$6 in Fe using Artificial Neural Network (ANN) models to gain insight into the stability and mobility of mixed copper–vacancy clusters, which play a key role in copper precipitation in iron alloys under irradiation. Castin~\etal also used ANNs  to predict vacancy migration energies as functions of the local atomic environment in Fe-based alloys~\cite{castin2010calculation,castin2018advanced}. Datasets of migration barriers for nearest-neighbour jumps on the Cu surfaces were calculated with the nudged elastic band (NEB) method and the tethering force approach of Kimari~\etal~\cite{kimari2020data} and used to train ANN models to predict the migration barriers for arbitrary nearest-neighbour Cu jumps. 
Fan \etal developed a convolutional neural network (CNN)—based model to predict the path-dependent vacancy migration energy barrier spectra in the TaNbMo multi-principal element alloy, using local chemical features surrounding each vacancy extracted via spatial density maps (SDMs)~\cite{fan2021predicting}.  Their framework can be applied to predict barriers in both crystalline materials~\cite{fan2021predicting} and amorphous materials~\cite{fan2022predicting2}. Lapointe \etal~\cite{lapointe2022machine} implemented a nonlinear surrogate model approach to learn and predict the kinetic transition rates during defect migration in $\alpha$-iron and amorphous Si systems. They show that it is possible to predict the formation entropy of defects and the logarithm of the prefactor of their activated events with only $O(N)$ numerical computations, which avoids the time-consuming evaluation of the system’s dynamical matrix $(O(N^2))$ and its spectrum $(O(N^3))$.
In their study of vacancy-mediated sluggish diffusion in concentrated Ni-Fe model alloys, Huang \etal~\cite{huang2023machine} used about 32,000 pre-calculated NEB~\cite{jonsson1998nudged, henkelman2000climbing} barriers to train an ANN-based machine learning model to predict the vacancy migration barriers for arbitrary local atomic environments, including both random solution configurations and alloys with short-range ordering.   

While these efforts have demonstrated the power of ML-guided kMC models to simulate defect evolution in complex alloys, none have considered whether the resulting kMC models obey key physical constraints. In particular, one important physical constraint is the so-called detailed balance condition (DBC). The DBC is a cornerstone of Monte Carlo methods since it provides a simple guarantee that an algorithm will sample from a well-defined stationary distribution in the long-time limit. While the DBC is a sufficient condition for the existence of a stationary distribution, it is not strictly necessary, as weaker conditions such as global balance can provide the same guarantees \cite{bernard2009event}. However, the DBC is typically much simpler to impose as it can naturally be enforced when transition rates are derived from Transition State Theory (TST).
 
In this paper, we investigate physics-informed generalizations of this type of ML approach, focusing on the imposition of the DBC. In the following, we mathematically demonstrate that the DBC-preserving structure of TST can be used to design ML architectures that are also guaranteed to exactly obey the DBC {\em by construction}. Using vacancy diffusion in concentrated alloys as an example,
we investigate the performance of different ML variants that attempt to impose the DBC using different non-strict strategies in order to identify possible trade-offs between accuracy of the prediction of individual rates and the extent to which the DBC is enforced. We find that architectures that strictly impose the DBC by construction in fact exhibit lower errors than their inexact counterparts, suggesting that the introduction of physically-motivated constraints in fact assists the learning process, providing higher-quality models that obey physical constraints without increasing the computational cost.

\section*{Theoretical Analysis}
\label{sec:theory}

We begin with a theoretical analysis of the DBC. In the following, quantities that are invariant to the exchange of the initial and final states are denoted with parenthesis subscripts, i.e., $f_{(A,B)} \equiv f_{(B,A)}$, while anti-symmetric quantities, i.e., those that only change sign upon exchange of the initial and final states are denoted by bracket subscripts, i.e., $f_{[A,B]} \equiv  - f_{[B,A]}$.

We first consider a class of kMC models where transition rates are assumed to follow from TST. In the case of an $A \rightarrow B$ transition, where $A$ and $B$ are global states of the system, the canonical transition rate can be written in term of the configurational partition function $Z_A$ of the initial state $A$ and of the configurational partition function of the dividing surface between states $A$ and $B$, denoted by $Z^*_{(A,B)}$, as: 

\begin{equation}
k_{A\rightarrow B} = \frac{1}{2}\sqrt{\frac{2}{\pi\beta}} \frac{Z^*_{(A,B)}}{Z_A} = \frac{\bar{Z}^*_{(A,B)}}{Z_A}
\label{eq:tst-fact}
\end{equation}
where $\beta = \frac{1}{kT}$, ($k$ is the Boltzmann constant) and the constants have been absorbed into the numerator in the last equation, simplifying the expressions. Note that since $Z^*_{(A,B)}$ corresponds to an integral over the (hyper)-surface between states $A$ and $B$, it is by construction invariant to the exchange of the initial and final states. TST rates are often further simplified by invoking the so-called harmonic approximation  of TST (HTST) \cite{vineyard1957frequency}, where the transition rates become:

\begin{equation}
  k_{A\rightarrow B} = \nu_{A\rightarrow B} e^{-\Delta E^b_{A \rightarrow B}/k_BT} = \nu_{A\rightarrow B} e^{-\beta(E^S_{(A,B)}-E_A)} = \frac{e^{-\beta E^S_{(A,B)}}}{\nu_{(A,B)}} \frac{\nu_A}{e^{-\beta E_A}}, 
  \label{eq:htransition_rate}
\end{equation}
where $\Delta E^b_{A \rightarrow B}$ is the energy barrier for the $A\rightarrow B$ transition, $E^S_{(A,B)}$ is the energy of the corresponding saddle point (S) and $E_A$ the energy of the initial state, $\nu_A$ and $\nu_{(A,B)}$ are the components of the vibrational prefactor for the state A and the saddle plane respectively . The notation  $E^S_{(A,B)}$ reflects the fact that the energy of the saddle point is the same for the forward and backward transitions (note that the forward and backward barriers are however in general not equal, see below). Following the same arguments as above, the component of the vibrational prefactor that pertains to the saddle plane $\nu_{(A,B)}$ is also the same in both directions. It then follows that such rates can also be written as $k_{A\rightarrow B} = \frac{\bar{Z}^*_{(A,B)}}{Z_A}$. That is, the rate can be expressed as the ratio of partition functions that depend on a common property between states $A$ and $B$ and one that depends only on state $A$.

\subsection*{Detailed balance condition}
\label{subsec:dbc}

Expressing transition rates in such a way is extremely beneficial as it provably 
leads to reversible Markov Chains that exactly obey the DBC \cite{kijima2013markov}. Indeed, the Kolmogorov criterion  \cite{kelly2011reversibility} is a necessary and sufficient condition for the DBC to hold. In the context of continuous-time Markov Chains, the criterion states that the product of the transition rates along {\em any} finite closed path should be independent of the direction in which the path is traversed, i.e.,
\begin{equation}
    k_{j_1\rightarrow j_2} k_{j_2\rightarrow j_3} ... k_{j_{n-1}\rightarrow j_{n}},k_{j_n\rightarrow j_1} = k_{j_1\rightarrow j_n} k_{j_{n}\rightarrow j_{n-1}}...k_{j_3\rightarrow j_2} k_{j_2\rightarrow j_1} 
\end{equation}
for all finite sequences of states. This equality is enforced by construction for transition rates of the form Eq.\ \ref{eq:tst-fact}, since 

\begin{equation}
 \frac{\bar{Z}^*_{(j_1,j_2)} \bar{Z}^*_{(j_2,j_3)} ... \bar{Z}^*_{(j_{n-1},j_{n})} \bar{Z}^*_{(j_{n},j_{1})}   }{ Z_{j_1} Z_{j_2} ... Z_{j_{n-1}} Z_{j_n} } \equiv \frac{\bar{Z}^*_{(j_1, j_n)} \bar{Z}^*_{(j_{n},j_{n-1})} ... \bar{Z}^*_{(j_3,j_2)} \bar{Z}^*_{(j_2,j_1)}   }{ Z_{j_1} Z_{j_n} ...  Z_{j_{3}} Z_{j_{2}}}, 
\end{equation}
 directly follows from $\bar{Z}^*_{(i,j)}\equiv \bar{Z}^*_{(j,i)}$. 

\subsection*{Consequences for ML approximations of transition rates}

The previous discussion suggests a simple avenue to enforce the DBC in machine-learned estimations by factoring the learning task into two sub-components following Eq.\ \ref{eq:tst-fact}. Note that we refer to these components as partition functions, by analogy with TST, but these can be seen as purely abstract trainable functions in a general ML setting.

Following the discussion above, these two components are: 
\begin{itemize}
    \item The partition function $Z_{A}$ corresponding to an initial state $A$. This term should take as input a featurization of a single state $A$ that is agnostic to specific final states of  transitions leading out of the state (i.e., $Z_{A}$ should be a state-wise quantity, not a transition-wise quantity). 
    \item The partition function $\bar{Z}^*_{(A,B)}$ corresponding to a specific transition between two states $A$ and $B$. This term can be computed using a featurization of both states $A$ and $B$ and/or of the transition state/transition path between $A$ and $B$. Crucially, both the {\em featurization and architecture} used to learn this term should be invariant with respect to the exchange of initial and final states $A$ and $B$ {\em by construction}.
\end{itemize}
These two terms can be combined to produce a transition rate for $A\rightarrow B$ transitions for any possible final states $B$ using Eq. \ref{eq:tst-fact},  which automatically guarantees that the DBC will be exactly obeyed.

\subsection*{Local approximations to the transition rates}
\label{subsec:locality}

Strictly speaking, the partition functions that enter the transition rates are global quantities that correspond to integrals over the full $3N_\mathrm{atoms}$ or $3N_\mathrm{atoms}-1$-dimensional configuration space, in the case of $Z_A$ and $\bar{Z}^*_{(A,B)}$, respectively. 
It can however formally be shown that very accurate local approximations can be developed in cases where  transitions are {\em spatially localized}, e.g., when unstable transition modes
are concentrated over a small subspace $\Omega^t$ of the whole configuration space $\Omega$ (c.f., Sects. 4 and 5 in Ref.\ \cite{binder2015analysis}). In this case, the partition functions can be expressed as integrals over $\Omega^t$ alone, albeit using an effective Hamiltonian than accounts for the interactions between resolved atoms in $\Omega_t$ and unresolved atoms in the complement set outside of the local domain $\Omega \setminus \Omega^t$. The analysis presented in Ref.\ \cite{binder2015analysis} formalizes the intuition that the details of the far-field atomic configuration away from the transition region should not affect transition rates, so that local information in the transition region should be sufficient to accurately approximate transition rates. (Note however that counter-examples to this local behavior do exist, e.g., when  transition pathways strongly couple with long-range displacement fields, e.g., through elastic interactions \cite{bagchi2024anomalous}, which can lead to extremely delocalized unstable modes.) In the context of ML, this indicates that it should be possible and desirable to learn transition rates using {\em local} featurizations around the transition region (e.g., where a defect is located) instead of using {\em global} featurizations of the whole configuration.

\subsection*{DBC and the composition of local approximations to the transition rates}
\label{subsec:local-dbc}
We now show that it is still possible to design ML approximations to the transition rates that obey the DBC even if more than one reactive local environment (i.e. akin to the subspace $\Omega^t$ defined above) is present in the same global configuration. That is, we can still construct ML models that satisfying the DBC when more than one defective region exists in the system. First consider an ensemble of $N$ disjoint local environments  embedded in a given atomic configuration. 
The joint state of the combined system can be indexed by $N$ sub-indices $\{\alpha_1,...,\alpha_N \}$, one for each local environment. The Markov Chain of the combined system can be made to obey the DBC if the joint effective partition functions are taken to factor as:

\begin{equation}
Z_{\{\alpha_1,...,\alpha_N \}} = \prod^N_{i=1} Z_{\alpha_i},
\end{equation}
and 
\begin{equation}
\bar{Z}_{(\{\alpha_1,...,\alpha_{j},...,\alpha_N \} , \{\alpha_1,...,\alpha'_{j},...,\alpha_N \} )} = \bar{Z}_{(\alpha_{j},\alpha'_{j})} \prod^{N, i\neq j}_{i=1}  Z_{\alpha_i}. 
\end{equation}

Evoking that of non-interacting systems, the partition function factorizations for a transition between local configuration 
$\alpha_j$ and $\alpha'_j$ in environment $j$ lead to transitions rates of the form 
\begin{equation}
k_{\{\alpha_1,...,\alpha_{j},...,\alpha_N \} \rightarrow \{\alpha_1,...,\alpha'_{j},...,\alpha_N \} } = \frac{ \bar{Z}_{(\{\alpha_1,...,\alpha_{j},...,\alpha_N \} , \{\alpha_1,...,\alpha'_{j},...,\alpha_N \} )}}{Z_{\{\alpha_1,...,\alpha_N \}} }
= \frac{\bar{Z}_{(\alpha_{j},\alpha'_{j})}}{Z_{\alpha_{j}}} = 
k_{\alpha_{j}\rightarrow \alpha'_{j}} 
\end{equation}
which exactly correspond to the local transition rate expression introduced above.
Direct inspection suffices to show that transition rates still factor into a ratio of a partition function that is invariant to the exchange of the initial and final states and of a partition function that depends only on the properties of the initial state, which is sufficient for the DBC to hold.

Now consider a case where some of the $N$ local environments do overlap with each other (say environments $i,j$ and $k$). In this case, a single transition can 
affect multiple local environments. If we enforce that each transition and its inverse can be unambiguously assigned to {\em the same} local environment (marked with a tilde symbol), the factorization proposed above, which now takes the form: 

\begin{equation}
\bar{Z}_{(\{\alpha_1,...,\alpha_{i},\tilde{\alpha_{j}},\alpha_{k},...,\alpha_N \} , \{\alpha_1,...,\alpha'_{i},\tilde{\alpha_{j}}',\alpha'_{k},...,\alpha_N \} )} = \bar{Z}_{(\tilde{\alpha_{j}},\tilde{\alpha_{j}}')} \prod^{N, h\neq j}_{h=1}  Z_{\alpha_h},
\end{equation}
still results in a kMC model that also exactly obeys the DBC, although the intuitive interpretation of the factorization in terms of partition functions of non-interacting sub-domains no longer holds. The same reasoning applies when transitions lead to the splitting or merging of local-environments, although care must then be taken to define domains and local features  consistently for the forward and backward jumps, which in general can be a difficult problem. Alternatively, the system could be partitioned into coarser non-overlapping "semi-local" environments where this difficulty could potentially be circumvented.

\subsection*{Specialization to energy barrier prediction}

\begin{figure}[ht!]
  \centering
  \includegraphics[width=4in]{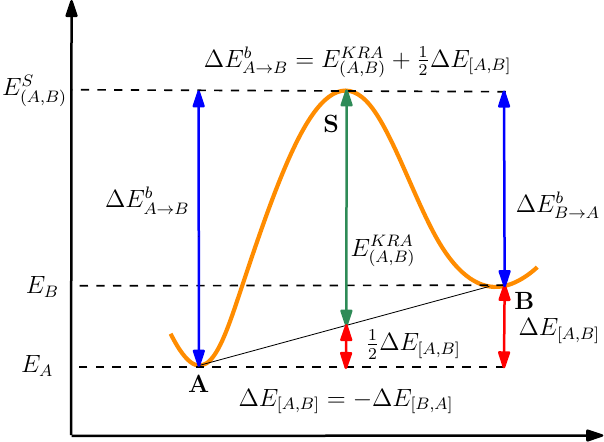}
  \caption {Schematic illustration of the transition between an initial state $A$ and final state $B$. The energy barrier for this transition can be decoupled into a thermodynamic component $\Delta E_{[A,B]} = E_B - E_A$ and a kinetic component given by the kinetically resolved activation barrier $E^{KRA}_{(A,B)}$.}
  \label{fig:energy_schematic}
  \end{figure}

In the following, we will consider the simplified problem of predicting
HTST rates using a constant "standard" state-independent prefactor $\nu$, a task which is equivalent to the prediction of energy barriers.
Generalization to rates of the form given by Eq.~\ref{eq:tst-fact} is conceptually straightforward.
As shown in Fig.~\ref{fig:energy_schematic}, this energy barrier can be decoupled into a thermodynamic component $\Delta E_{A,B} = E_B - E_A$ and a kinetic component given by the kinetically resolved activation barrier $\Delta E_{KRA}$ defined as:
\begin{equation}
\Delta E^{b}_{A\rightarrow B} = \Delta E^{KRA}_{A,B} + \frac{1}{2} \Delta E_{A,B}
\label{eqn:migration_energy}
\end{equation}
which, according to Eq.\ \ref{eq:htransition_rate}, yields transition rates of the form:
\begin{equation}
    k_{A\rightarrow B} = \nu \exp\left(-\beta \left[\Delta E^{KRA}_{A,B} +\frac{1}{2}\Delta E_{B,A} \right] \right)
    =  \frac{\nu \exp\left(-\beta \left[\Delta E^{KRA}_{A,B} +\frac{1}{2}\Delta E_{B,A}  + \mathcal{C}_A \right] \right)}{\exp\left(-\beta  \mathcal{C}_A \right) }
    \label{eq::ml_rate}
\end{equation}
where the last term on the RHS follows by multiplying by $1=\exp\left(-\beta  \mathcal{C}_A \right)/\exp\left(-\beta  \mathcal{C}_A \right)$ for some (yet undetermined) constant $\mathcal{C}_A$.

The last form of Eq.\ \ref{eq::ml_rate} indicates that the DBC will be obeyed so long as a {\em state-specific} constant $\mathcal{C}_A$ (i.e., that would numerically take the same value in expressions for $k_{A\rightarrow C}$ for any final states $C$) can be defined so as to render 
the numerator strictly invariant to the exchange of the initial and final states. If this can be achieved, rates can be expressed in the general form given by Eq.\ \ref{eq:tst-fact}, which by construction obeys the DBC. The requirement for the DBC can also be re-expressed in the context of the Kolmogorov criterion, which for any closed path ${j_1,j_2,...,j_{n-1},j_n, j_1}$ reduces to  
\begin{equation}
    \Delta E_{j_2,j_1} + \Delta E_{j_3,j_2} + ...  + \Delta E_{j_{n},j_{n-1}}+ \Delta E_{j_1,j_n}  = \Delta E_{j_n,j_1} + \Delta E_{j_{n-1},j_n} + ... + \Delta E_{j_2,j_3} + \Delta E_{j_1,j_2},
    \label{eq:kolmogorov_delta}
\end{equation}
so long as $\Delta E^{KRA}_{A,B}\equiv \Delta E^{KRA}_{(A,B)}$ is learned in a way that is strictly invariant to the exchange of $A$ and $B$. 
This equality enforces that the sum of the $\Delta E$  along a closed path should be independent of the direction in which it is traversed. Under a strict physical interpretation in terms of total potential energy differences, $\Delta E_{A,B} \equiv \Delta E_{[A,B]}$, and so the only physical solution to Eq.\ \ref{eq:kolmogorov_delta} is when each side of the equality is 0. It is easy to see that both sides of the equality Eq.\ \ref{eq:kolmogorov_delta} become equal to zero if $\Delta E_{[A,B]}= \mathcal{E}_B-\mathcal{E}_A$ where $\mathcal{E}_i$ is a function of state $i$ alone. Choosing $\mathcal{C}_A=\mathcal{E}_A$ then leads to rates of the form
\begin{equation}
    k_{A\rightarrow B} = 
      \frac{\nu \exp\left(-\beta \left[\Delta E^{KRA}_{(A,B)} +\frac{1}{2}(\mathcal{E}_B + \mathcal{E}_A) ) \right] \right)}{\exp\left(-\beta  \mathcal{E}_A \right) } ,
\label{eq:barrier}
\end{equation}
which obey the DBC by construction since the numerator is invariant to exchange of states $A$ and $B$ and the denominator depends only on the initial state $A$. 

We stress that enforcing strict anti-symmetry of the energy differences, i.e., $\Delta E_{A,B} \equiv \Delta E_{[A,B]}$, does not in itself guarantee that the DBC will be obeyed, as it only shows that the Kolmogorov criterion is obeyed by length 2 cycles, which is necessary but not sufficient for the DBC to hold in general. As discussed above, the Kolmogorov criterion should hold for arbitrary length of cycles in order to satisfy the DBC.

To summarize, the following two conditions are sufficient to ensure that energy barriers predicted by ML lead to transition rates that obey the DBC when used in conjunction with TST with a standard prefactor (Eq.\ref{eq:barrier}):
\begin{itemize}
    \item The thermodynamic energy difference $\Delta E_{[A,B]}$ should be expressed as $\mathcal{E}_B-\mathcal{E}_A$, a difference of {\em state-wise} constants. The ML architectures used to learn each term should take as input a featurization of the corresponding
state that is agnostic to possible final states of transitions leading out of the state.
\item The kinetically resolved activation energy $E^{KRA}_{(A,B)}$ should be computed
using a featurization of both states $A$ and $B$ and/or of the transition state/transition path between $A$ and $B$. Crucially, both the featurization and architecture used to learn this term should be invariant with respect to the exchange of initial and final states $A$ and $B$ by construction.
\end{itemize}

\section*{Results}
We now turn to constructing ML models that incorporate the DBC to various levels of approximation, to test the ability of different approaches to satisfy the DBC and to determine the impact of introducing additional terms to their respective loss functions on the overall quality of the model. We first present a ML approach that exactly obeys the conditions derived above and demonstrate its performance on the problem of predicting mono-vacancy jump kinetics in a CuNi binary alloy.  Around 30,000 barriers were generated for three average compositions. The models were trained on 75\% of this data ($D_{train}$) while the validation ($D_{val}$) and test sets ($D_{test}$) comprised the remaining 15\% and 10\% respectively. The barriers were computed using the LAMMPS molecular dynamics code~\cite{lammps}. The details of the data generation process are reported in the Methods section. 

Following the theoretical derivations above, the barrier prediction is first partitioned into two sub-problems: the prediction of thermodynamic energy differences  and of kinetically-resolved activation barriers, respectively.

\subsection*{Thermodynamic energy difference ($\Delta E$)}
\label{sec:de_dbc}

According to the prescription derived above, the energy difference between any two states $A$ and $B$ should be expressed as $\Delta E_{[A,B]}= \mathcal{E}_B-\mathcal{E}_A$, where each term is a state-wise constant inferred from a featurization that is agnostic to possible final states of the vacancy. To so characterize local environments, we use what we refer to as \textit{Kolmogorov} fingerprints.

To generate these fingerprints for any two arbitrary states $A$ and $B$, we first isolate atomic environments centered at the respective location of the vacancy using a cutoff radius of 6 \r{A} (this choice will be discussed in the Methods section).  Atomic environment vectors are then generated for these initial and final environments with respect to the vacancy site, following the procedure described in the Methods section. The vectors for each state are then concatenated to generate the two global fingerprints $\FGA$ and $\FGB$ corresponding to the initial state $A$ and the final state $B$ respectively. 

\begin{figure}[!ht]
\centering
\includegraphics[width=4in]{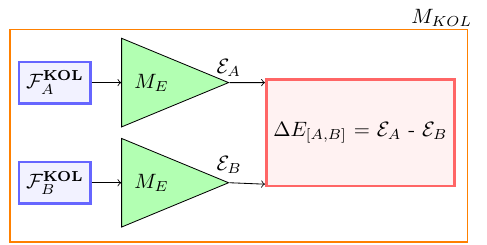}
\caption {Architecture of the DBC-obeying model $M_{KOL}$ that predicts the thermodynamic energy difference $\Delta E$ between a pair of states $A$ and $B$. See text for details. }
\label{fig:Kolmogorov_model}
\end{figure}

These fingerprints are then input to an ML architecture that obeys $E_{[A,B]}\equiv -E_{[B,A]}$, which we refer to as a 
 $Kolmogorov$-constrained model ($M_{KOL}$). As illustrated in Fig.~\ref{fig:Kolmogorov_model}, the \textit{Kolmogorov} fingerprints $\FGA$ and $\FGB$ of both the initial and final states are each fed to {\em identical} copies of a DNN model $M_{E}$. The respective outputs of these two DNN are then subtracted to yield a prediction of the thermodynamic energy difference $\Delta E_{[A,B]}= \mathcal{E}_B-\mathcal{E}_A$. This architecture enforces the anti-symmetry of the thermodynamic energy differences, since, when the inputs to $M_{E}$ are reversed, the output of $M_{KOL}$ takes the form $\mathcal{E}_A-\mathcal{E}_B$. We emphasise that this local approach relies on a featurization of only immediate neighborhood of the vacancy. The energies $\mathcal{E}_A$ and $\mathcal{E}_B$ should therefore not be interpreted as approximations of the total energy of the system, but as effective intermediate quantities introduced for the purpose of estimating energy differences between initial and final states of a transition, but which are still state-wise in $A$ and $B$.
 
\begin{figure}[h!]
    \begin{subfigure}[t]{0.33\textwidth}
      \includegraphics[width=\textwidth]{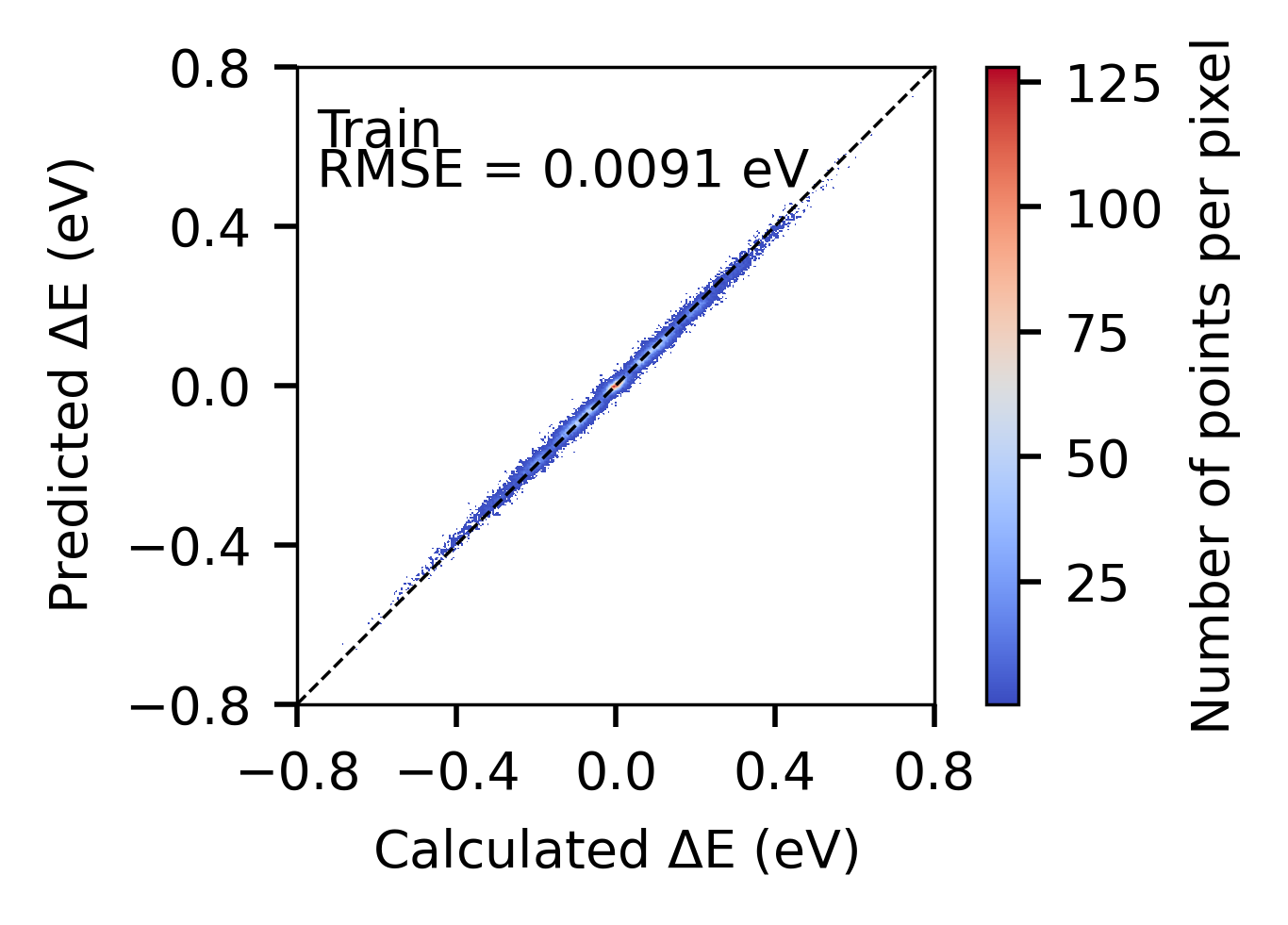}
      \caption{$\Delta_E$: Train}
    \end{subfigure}
    \hfill
    \begin{subfigure}[t]{0.33\textwidth}
      \includegraphics[width=\textwidth]{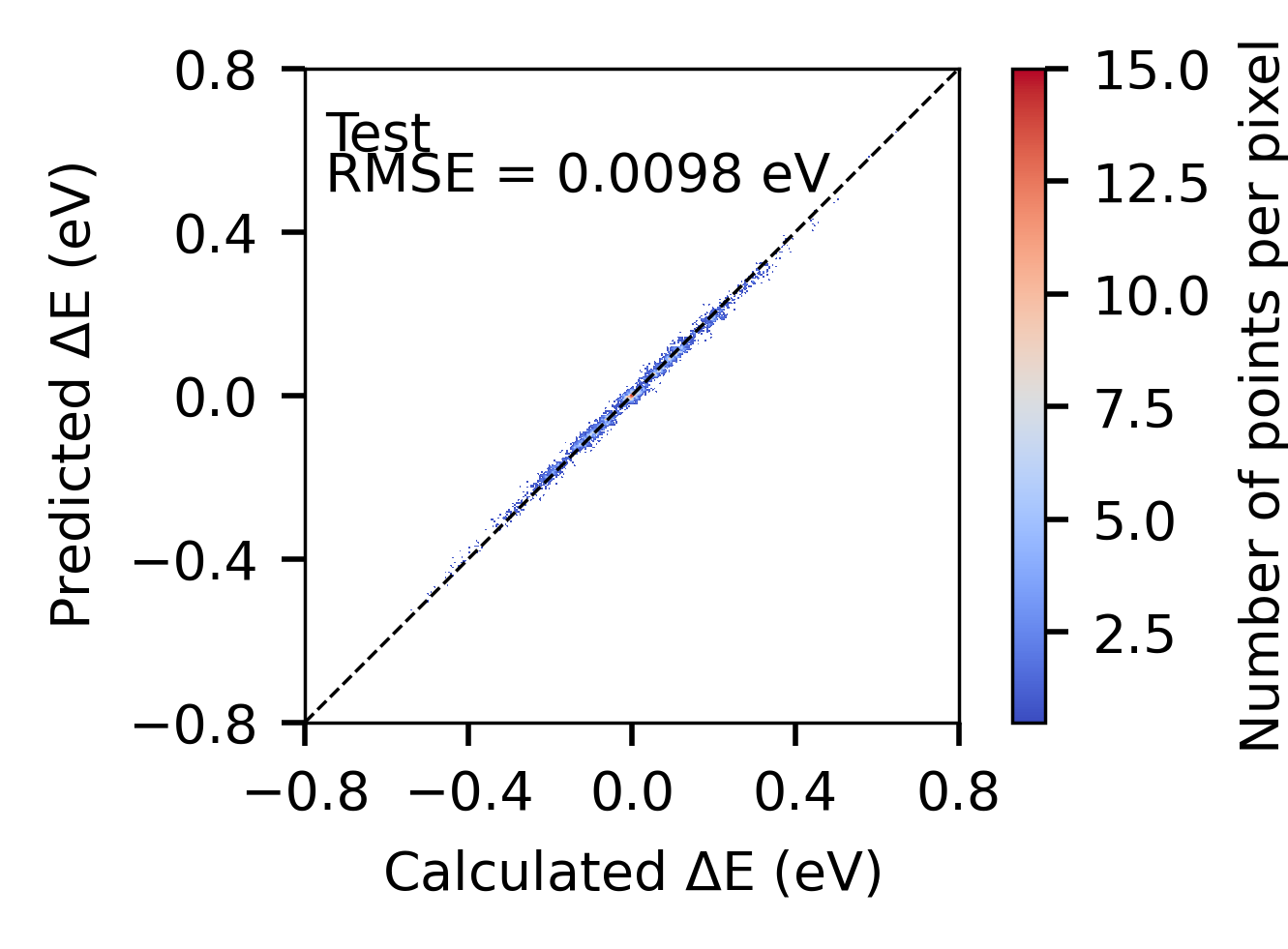}
      \caption{$\Delta_E$: Test}
    \end{subfigure}
    \begin{subfigure}[t]{0.33\textwidth}
      \includegraphics[width=\textwidth]{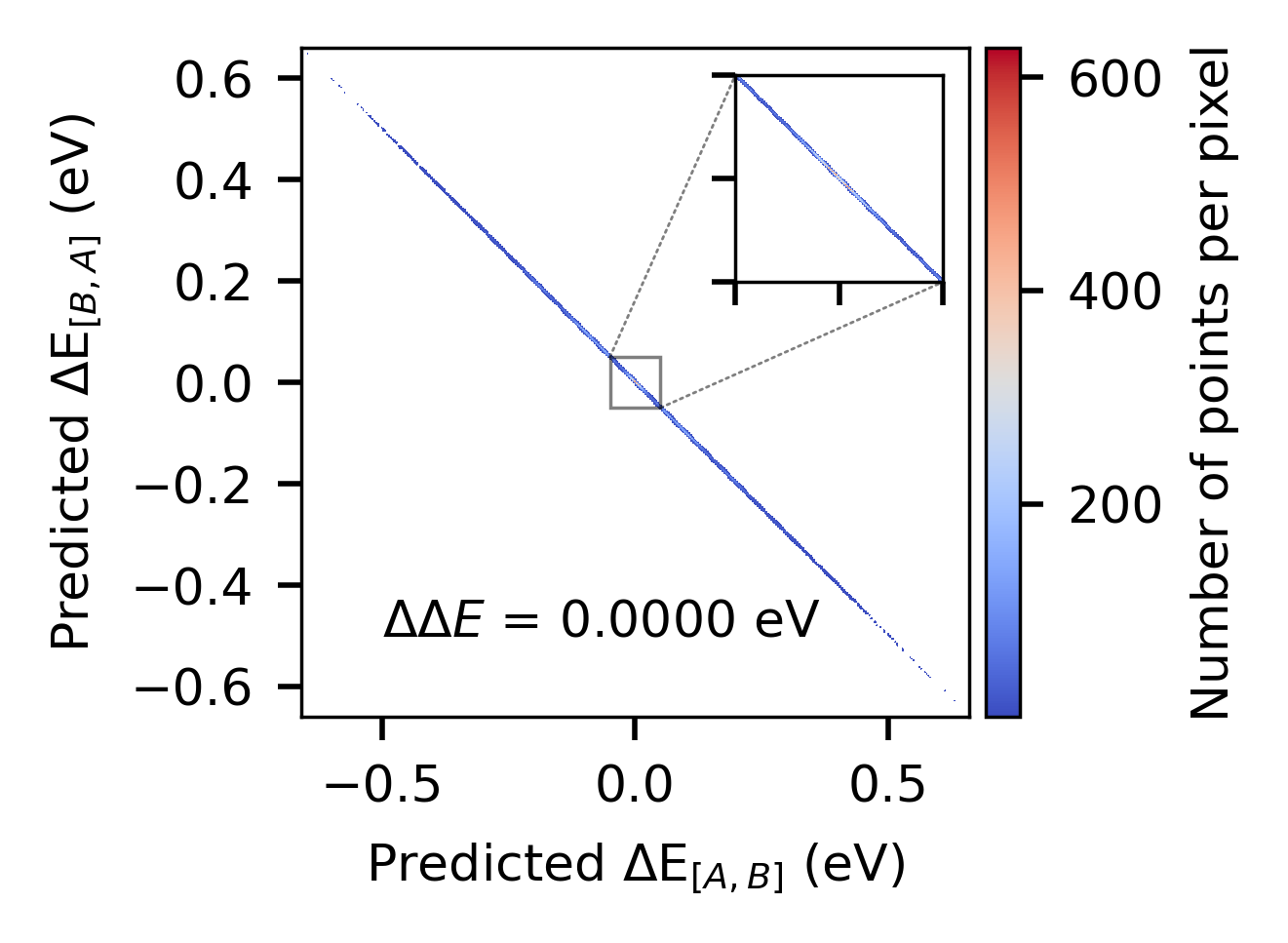}
      \caption{$\Delta \Delta_E$: Test}
    \end{subfigure}
    \caption{Results for the \textit{Kolmogorov}-constrained model $M_{KOL}$ for the prediction of thermodynamic energy differences $\Delta E$. (a) $M_{KOL}$ predictions on training data, b)  $M_{KOL}$ predictions on test data and c) Parity plot comparing the $\Delta E$ predictions in the forward direction ($\Delta E_{[B,A]}$) on the X-axis and the reverse direction ($\Delta E_{[A,B]}$) along the Y-axis. The perfect anti-symmetry of the prediction is shown in the inset.}
\label{fig:GHC_results}
  \end{figure}
  
$M_{KOL}$ is trained by minimizing the objective function $f_{min}^{KOL}$ :
\begin{equation}
\label{KOL_objective_fn}
f_{min}^{KOL} = \sqrt{\frac{1}{N}\sum_{i=1}^{N}(\Delta E_{ML}^i-\Delta E_{C}^i)^{2}},
\end{equation}
where $\Delta E_{ML}$ is the prediction by $M_{KOL}$ for the energy difference between a pair of arbitary states $A$ and $B$ and 
 $\Delta E_{C}$ is the corresponding ground truth value, i.e, the calculated energy difference between a pair of states 

There are effectively two performance metrics by which $M_{KOL}$ can be assessed. The first quantifies how well $M_{KOL}$ actually predicts $\Delta E$ and the second the extent to which the predictions obey the Kolmogorov criterion. Note that a model could perfectly obey detailed balance but do poorly in predicting $\Delta E$ and vice versa, although perfect predictions of $\Delta E$ would imply that the Kolmogorov condition is satisfied.
The performance of the model in maintaining detailed balance is denoted by a root mean square error (RMSE) $\Delta \Delta E$, which is defined as:
\begin{equation}
\Delta \Delta E = \sqrt{\dfrac{1}{N}\sum_{i=1}^{N} (\Delta E_{ML:[B,A]} + \Delta E_{ML:[A,B]})^2}
\end{equation}
where $\Delta E_{ML:[B,A]}$ and $\Delta E_{ML:[A,B]}$ are the $M_{KOL}$ predictions in the forward and reverse directions respectively for the test data $D_{test}$.

The training and test results are reported in Fig.~\ref{fig:GHC_results}a \& b respectively. Very low RMSE values of 0.0091 eV and 0.0098 eV are obtained on both the training and test data respectively, significantly lower than typical thermal energies. Fig.~\ref{fig:GHC_results}c confirms that $M_{KOL}$ predicts thermodynamic energy differences that are perfectly anti-symmetric with respect to the exchange of initial and final states (i.e., $\Delta \Delta E = 0 $), fulfilling a first key requirement for the DBC.  While $M_{KOL}$ does guarantee the DBC by construction, we will investigate another architecture later where $\Delta \Delta E$ is identically zero for all transitions, but nonetheless the DBC is not obeyed.

\label{sec:results}
\subsection*{Kinetically resolved activation barrier ($E^{KRA}_{(A,B)}$)}

\label{sec:kra_dbc}
The second step of the barrier prediction is that of the kinetically resolved activation barrier $E^{KRA}_{(A,B)}$. 
As described in the previous section, it should be computed using a featurization of both states $A$ and $B$ and/or of the transition state/transition path between $A$ and $B$. Further, both the featurization and architecture used to learn $E^{KRA}_{(A,B)}$ should by construction be invariant with respect to the exchange of initial and final states $A$ and $B$.

For this task, we introduce \textit{transition-wise} configurational environment fingerprints. To do so, local environments are first constructed around the target vacancy for the initial ($A$) and final ($B$) states, as above. In contrast to the prediction of energy differences, the union of the two sets of atoms is then constructed, forming a single transition-specific set that surrounds both the initial and final locations of the vacancy. The properties of this set of atom in both the initial and final states are then obtained as above, producing the two fingerprints $\FTWA$ and $\FTWB$ describing the initial and final environments around the vacancy. It then follows that, in the case of \textit{transition-wise} fingerprints, $A$ and $B$ are always neighboring states. These extended fingerprints are expected to be more efficient at allowing ML to learn the properties of the saddle point/dividing surface corresponding to the target transition compared to purely state-specific fingerprints. 

\begin{figure}[ht!]
    \begin{subfigure}[t]{0.3\textwidth}
      \includegraphics[width=\textwidth]{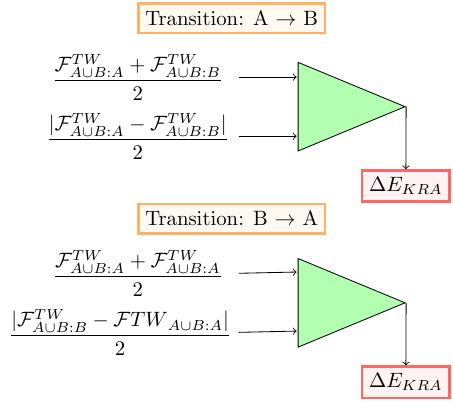}
      \caption{DNN schematic}
    \end{subfigure}
    \hfill
    \begin{subfigure}[t]{0.34\textwidth}
      \includegraphics[width=\textwidth]{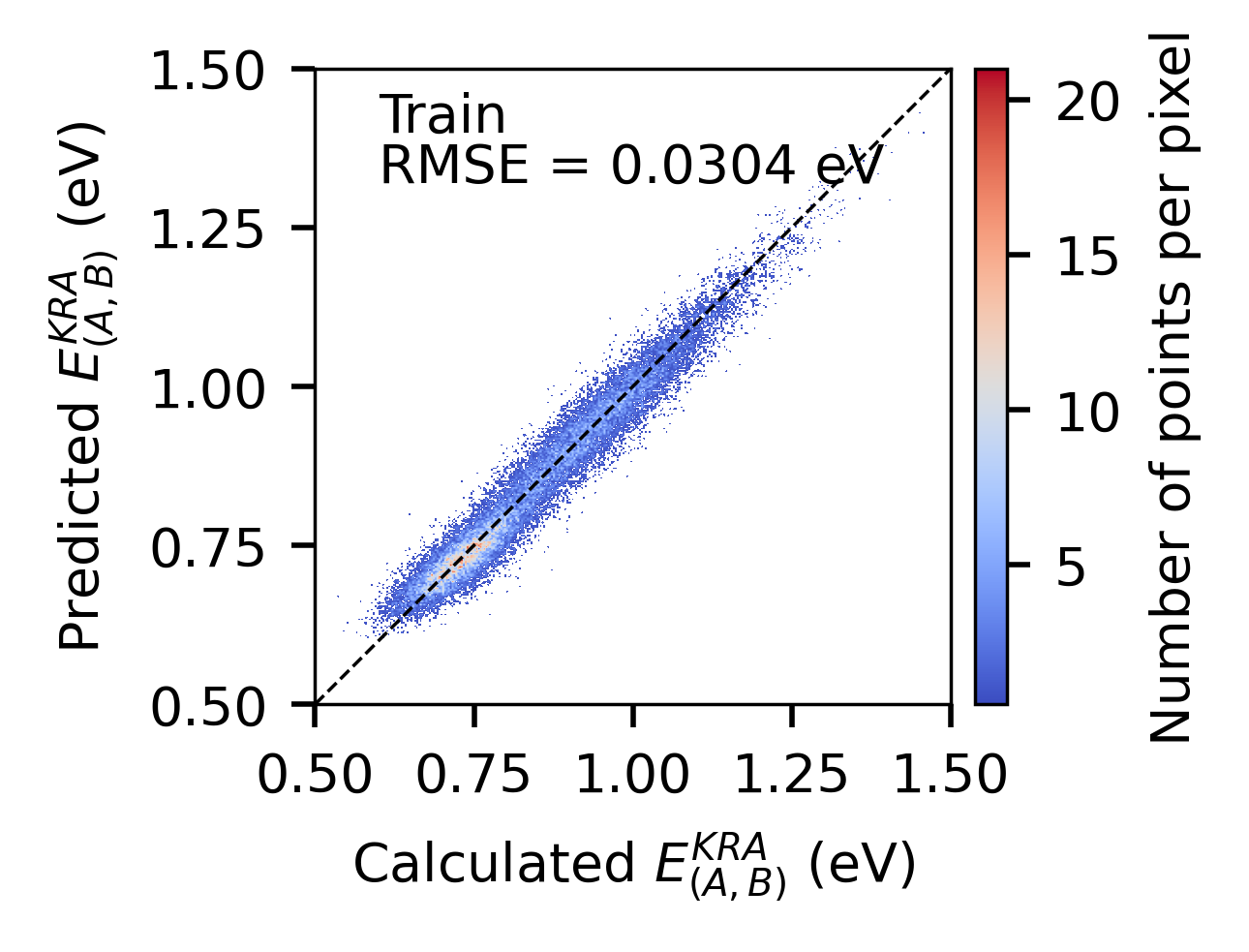}
      \caption{$E^{KRA}_{(A,B)}$ Train}
    \end{subfigure}
    \begin{subfigure}[t]{0.34\textwidth}
      \includegraphics[width=\textwidth]{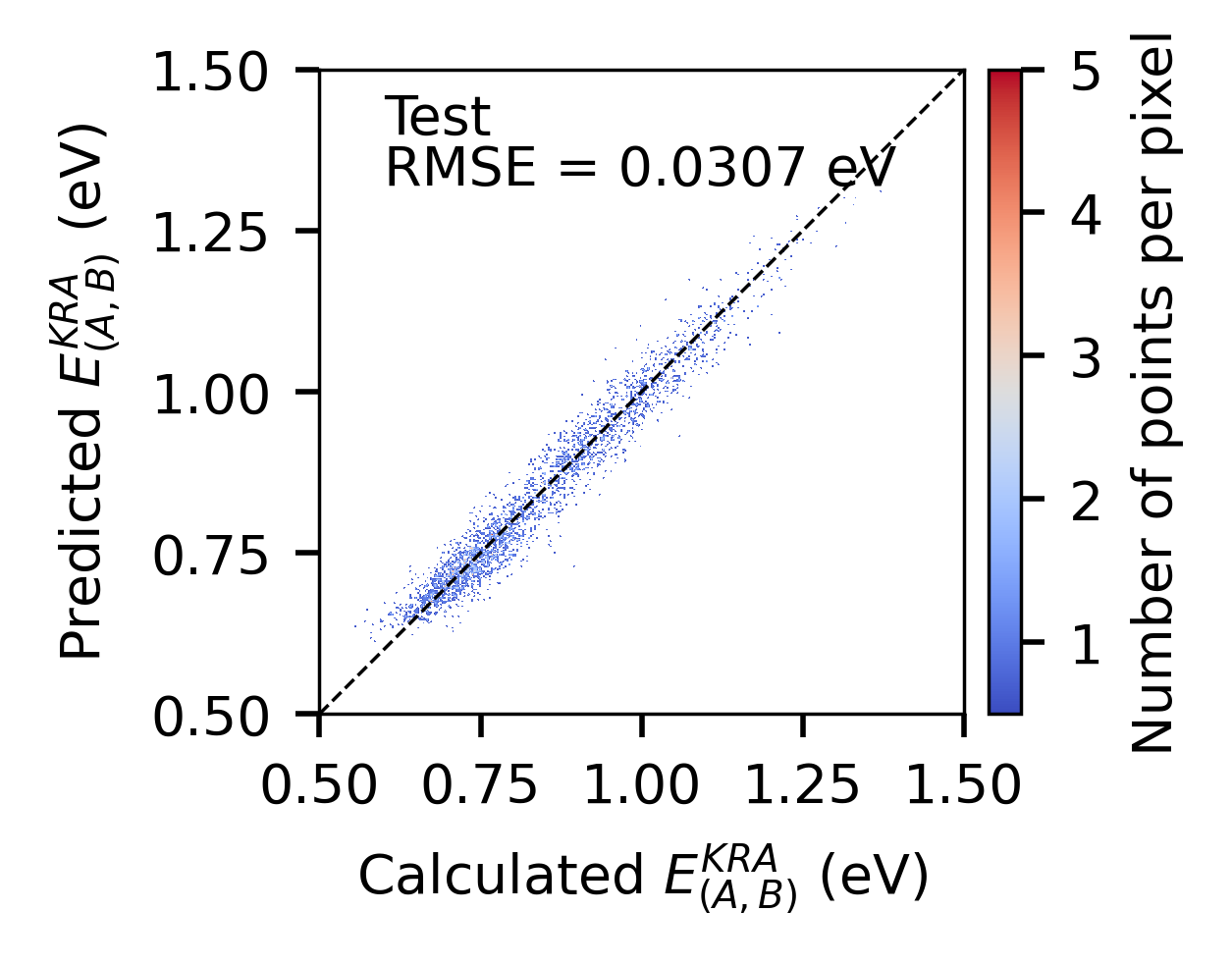}
      \caption{$E^{KRA}_{(A,B)}$ Test}
    \end{subfigure}
    \caption{DNN model and results for prediction of $E^{KRA}_{(A,B)}$. (a) Model architecture using weighted sums and differences of \textit{transition-wise} fingerprints to train model while maintaining symmetry constraints and b) training and testing parity plots for the model.}
\label{fig:E_kra_results}
  \end{figure}

The second key requirement is to ensure invariance of the prediction of $E^{KRA}_{(A,B)}$ with respect to exchange of the initial and final states. A simple strategy to enforce this 
is to symmetrize the input feature themselves by forming the sum $\dfrac{1}{2}(\FTWA +\FTWB)$  and absolute  difference $\dfrac{1}{2}|\FTWA-\FTWB|$ of the two \textit{transition-wise} fingerprints. 
These two features are then used as input to a DNN, as shown in Fig.~\ref{fig:E_kra_results}a. This ensures an output that is symmetric with respect to permutation of the initial and final states, since both situations would yield identical inputs. (Note that other strategies are possible, e.g., by replacing the final difference in the $M_{KOL}$ architecture by a permutation invariant function such as a sum or a product.) The performance of this model is reported in Fig.~\ref{fig:E_kra_results}b \& c. 
The average RMSE on the training and test sets were found to be 0.0304 eV and 0.0307 eV respectively.
These errors are generally larger than for $M_{KOL}$, but are still on the order of $k_B T$, and so are expected to produce very accurate kinetics. As discussed in the Methods section, these results could potentially be further improved by increasing the size of the atomic environments around the vacancies.

Overall, these results show that simple DBC-obeying architectures perform extremely well at predicting the energy barriers for vacancy hopping in a concentrated alloy, yielding errors on the same order as the thermal energy, which in general can be expected to be smaller than the error incurred by the reference method unto which the model is trained, be it either empirical potentials, ML potentials, or even approximate quantum methods such as DFT. Enforcing the DBC by construction therefore does not appear to lead to significant erros in prediction accuracy. In fact, as we now show through comparisons with three different non-DBC compliant ML implementations, it appears that enforcing the DBC actually {\em improves} the accuracy of the barrier prediction.
  
\subsection*{Traditional Non-DBC compliant implementation}
In the following, we consider three ML variants that attempt at imposing the DBC for the thermodynamic energy difference $\Delta E$ using approximate strategies in an attempt to quantify possible 
tradeoffs between accuracy of the prediction of individual rates and the extent to which the DBC is enforced. 

A possible origin of this potential tradeoff is the use of {\em transition-agnostic, state-wise} features to estimate thermodynamic energy differences. Indeed, one could expect that, just like transition-wise features can be expected to be more accurate at capturing the properties of the saddle point between two states, they could also improve the estimation of the energy difference between two neighboring states. To assess this possibility, we consider three separate variants of energy difference estimation using such transition-wise features, namely: i) a \textit{no-constraint} model ($M_{NC}$), ii) a \textit{soft anti-symmetry constraint} model ($M_{SAS}$) and iii) a \textit{hard anti-symmetry constraint} model ($M_{HAS}$).  The model details are summarized in Table~\ref{tab:detailed_balance} and will be discussed in detail below. 

In all of these implementations, the input consists of transition-wise features computed using the procedure described earlier for the prediction of $E^{KRA}_{(A,B)}$. However, once extracted, the transition-wise fingerprints  $\FTWA$ and $\FTWB$ are simply concatenated into a single long vector, and used directly.

\begin{flushleft}
\begin{table}
\centering
\begin{tabular}{|m{2.1cm}|m{7.2cm}|m{3.6cm}|m{2cm}|}
\hline
Model & \centering{DNN Architecture} & \centered{Objective\\ Function ($f_{min}$)}  & Description \\ \hline
\centered{No-\\ constraint  \\ ($M_{NC}$)}    & \includegraphics[height=3.5cm]{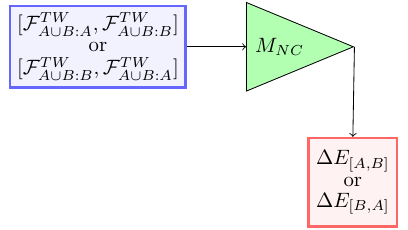}   & \centered{ $\sqrt{\frac{1}{N}\sum_{i=1}^{N}(\Delta E_{ML}^i-\Delta E_{C}^i)^2}$}  & \centered{Model learns \\ from only \\ data}\\ \hline

\centered{Soft \\ anti-symmetry\\ constraint  \\ ($M_{SAS}$)}   & \includegraphics[height=3.5cm]{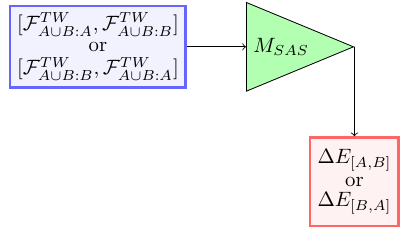}  &  {$\begin{aligned} & \sqrt{\frac{1}{N}\sum_{i=1}^{N}(\Delta E_{ML}^i-\Delta E_{C}^i)^2} \\ &+ \\  & \lambda \frac{1}{N}\sum_{i=1}^{N}(|\Delta E_{[B,A]}^i-\Delta E_{[A,B]}^i|) \end{aligned}$} & \centered{Penalty term \\ is introduced \\ in objective \\ function to \\ enforce anti- \\ symmetry} \\ \hline
\centered{Hard \\anti-symmetry \\ constraint\\ ($M_{HAS}$)}      & \includegraphics[height=3.0cm]{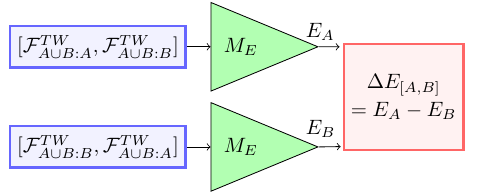}  & $\sqrt{\frac{1}{N}\sum_{i=1}^{N}(\Delta E_{ML}^i-\Delta E_{C}^i)^2}$ & \centered{Anti- \\ symmetry \\ constraint is \\ hard-wired \\ into model \\ architecture}  \\ \hline
\end{tabular}
\caption{Comparison of the DNN architecture and objective functions to be minimized for three non-strictly DBC-obeying  methods for the prediction of $\Delta E$.  All listed models use \textit{transition-wise} fingerprints $\FTWA$ and $\FTWB$, which indicates that $A$ and $B$ are neighboring states. See text for details.}
\label{tab:detailed_balance}
\end{table}
\end{flushleft}
 
\subsubsection*{\textit{No-constraint} model ($M_{NC}$)}
In the \textit{no-constraint} implementation, a single DNN $M_{NC}$ is trained without enforcing any additional constraint, which corresponds to the most commonly encountered approach in the literature. 

The objective function $f_{min}^{NC}$ that is minimized is the same that is used in $M_{KOL}$ and is defined as:
  \begin{equation}
  \label{NC_objective_fn}
  f_{min}^{NC} = \sqrt{\frac{1}{N}\sum_{i=1}^{N}(\Delta E_{ML}^i-\Delta E_{C}^i)^{2}},
  \end{equation}
where, $\Delta E_{ML}$ is the prediction of the DNN model $M_{NC}$ and $\Delta E_{C}$ is the ground truth, i.e the calculated energy difference between the initial state $A^i$ and final state $B^i$ for each of the $N$ hops. The (forward or reverse) direction of the transition is implicitly encoded by the order in which the features of the initial and final states are concatenated. To predict $\Delta E_{[A,B]}$, the fingerprints are concatenated as $[\FTWA, \FTWB]$, while to predict $\Delta E_{[B,A]}$, the fingerprints are concatenated as $[\FTWB, \FTWA]$.
The schematic for this model architecture is presented in Table~\ref{tab:detailed_balance}.

\begin{figure}[h!]
    \begin{subfigure}[t]{0.33\textwidth}
      \includegraphics[width=\textwidth]{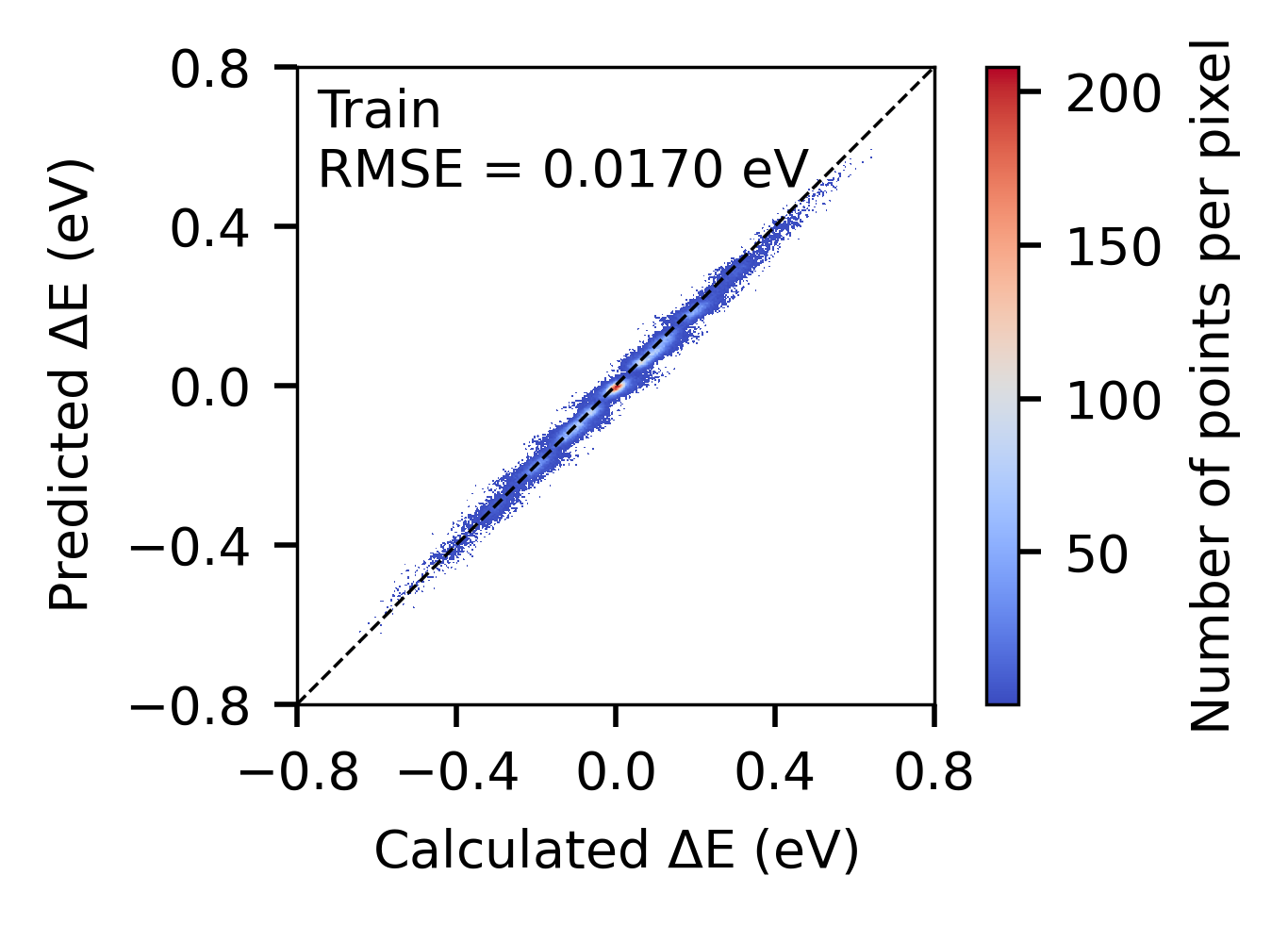}
      \caption{$\Delta_E$: Train}
    \end{subfigure}
    \hfill
    \begin{subfigure}[t]{0.33\textwidth}
      \includegraphics[width=\textwidth]{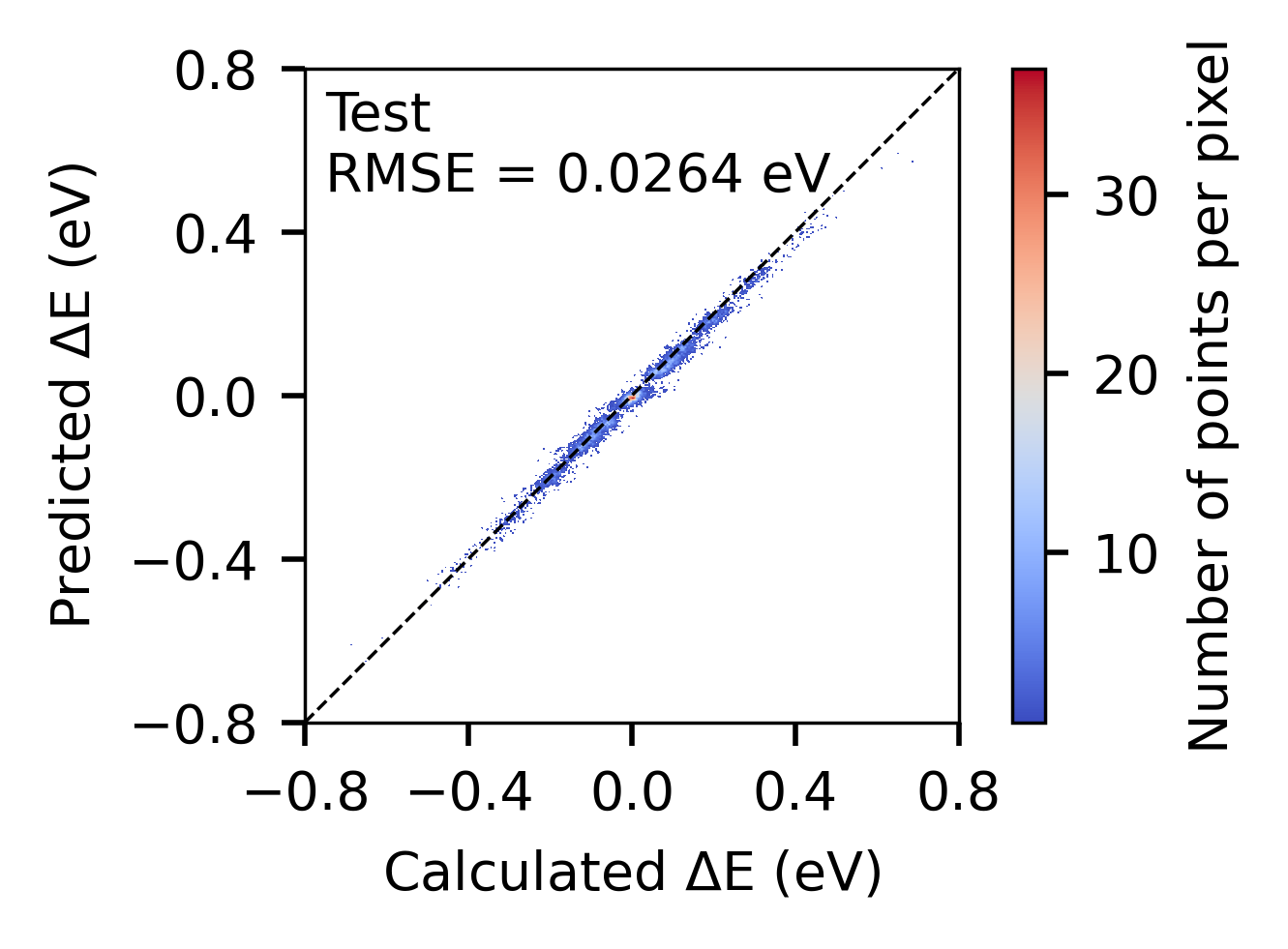}
      \caption{$\Delta_E$: Test}
    \end{subfigure}
    \begin{subfigure}[t]{0.33\textwidth}
      \includegraphics[width=\textwidth]{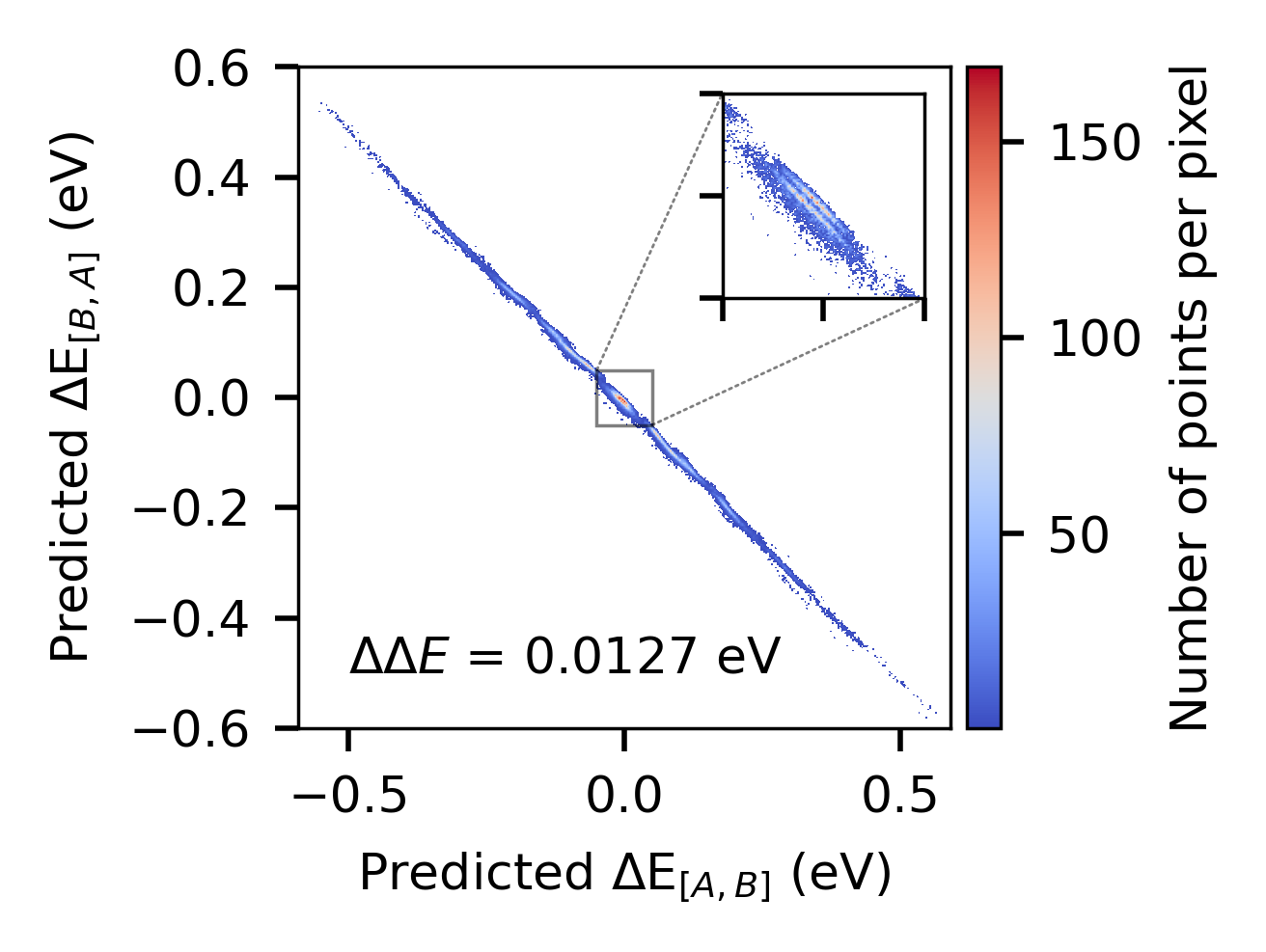}
      \caption{$\Delta \Delta_E$: Test}
    \end{subfigure}
    \caption{Performance of the \textit{no-constraint} $M_{NC}$ model to the prediction of the energy difference $\Delta E$. (a) $M_{NC}$ predictions on training data, b)  $M_{NC}$ predictions on test data and c) Parity plot comparing $\Delta E$ predictions in the forward direction ($\Delta E_{[A,B]}$) on the X-axis and the reverse direction ($\Delta E_{[B,A]}$) along the Y-axis. The inset shows the deviation from perfect anti-symmetry $\Delta \Delta E = 0.0127 eV$.}
\label{fig:NC_results}
  \end{figure}

Training and test results for $M_{NC}$ are reported in Fig.~\ref{fig:NC_results}a \& b respectively. RMSE values of 0.0170 eV  and 0.0264 eV are obtained on $D_{train}$ and $D_{test}$ respectively, indicating that this model does well, although slightly worse than $M_{KOL}$, in predicting the magnitude of $\Delta E$. Fig.~\ref{fig:NC_results}c shows the performance of the model in obeying strict anti-symmetry. We see that there is substantial difference between the predicted $\Delta E$ in the forward and reverse directions. An average value of 0.0127 eV for $\Delta \Delta E$ is obtained on $D_{test}$. While this might appear to be a small error, we will see below that such errors can dramatically compound. 

\subsubsection*{\textit{Soft anti-symmetry constraint} model ($M_{SAS}$)}

A simple approach to alleviating the anti-symmetry violations in $\Delta E$ is to introduce a "soft-constraint" by including an explicit penalty 
term in the loss function. The strength of this penalty is controlled by a factor $\lambda$, yielding a loss function of the form: 

\begin{equation}
\label{eqn:SC_objective_fn}
f_{min}^{SAS} = f_{min}^{NC} +  \lambda \frac{1}{N}\sum_{i=1}^{N}(|\Delta E_{[B,A]}^i-\Delta E_{[A,B]}^i|)
\end{equation}
where $\Delta E_{[B,A]}$  and $\Delta E_{[A,B]}$ are the energies predicted by the DNN $M_{SAS}$ in the forward direction $(A \rightarrow B)$ and 
the reverse direction $(B \rightarrow A)$ respectively. $\lambda$ was varied from $\lambda = 0$ to $\lambda = 5$ in 
Eqn.~\ref{eqn:SC_objective_fn},  $f_{min}^{SAS}$ is found to be minimized for $\lambda = 0.2$. 
The schematic for this model architecture is indicated in Table~\ref{tab:detailed_balance}. 
Note that when $\lambda = 0$, $M_{SAS}$ reduces to $M_{NC}$.

\begin{figure}[h!]
    \begin{subfigure}[t]{0.33\textwidth}
      \includegraphics[width=\textwidth]{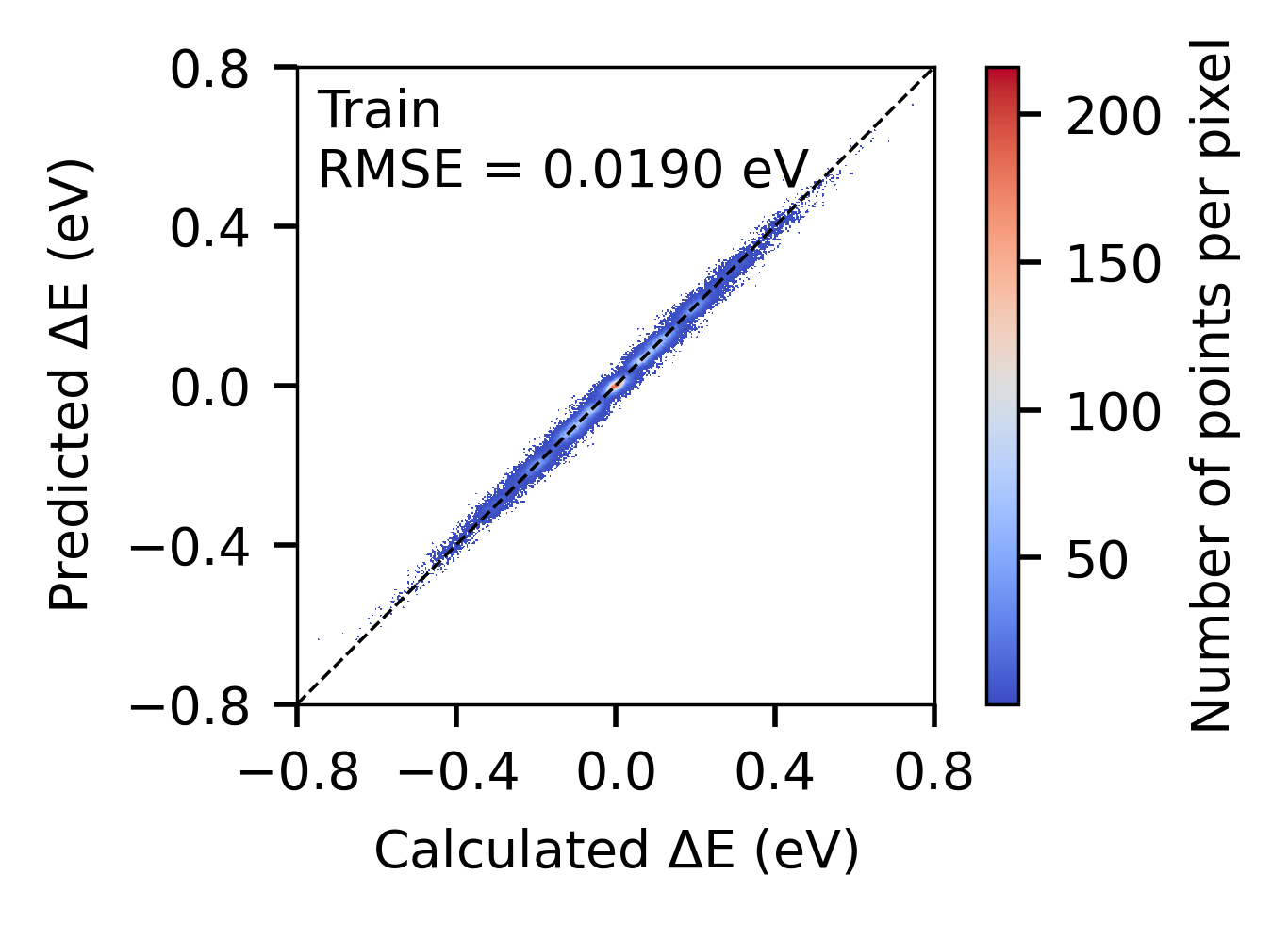}
      \caption{$\Delta_E$: Train}
    \end{subfigure}
    \hfill
    \begin{subfigure}[t]{0.33\textwidth}
      \includegraphics[width=\textwidth]{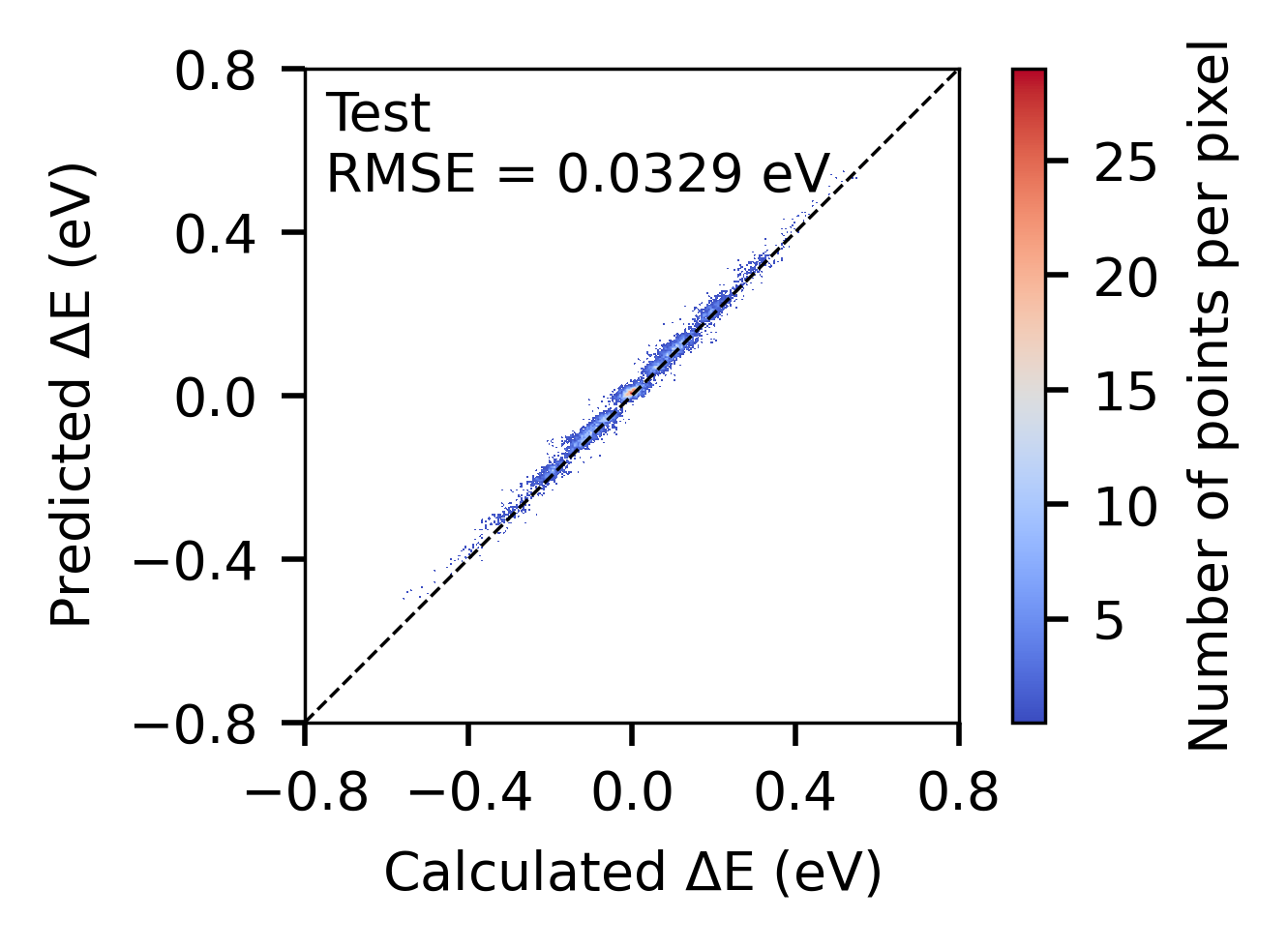}
      \caption{$\Delta_E$: Test}
    \end{subfigure}
    \begin{subfigure}[t]{0.33\textwidth}
      \includegraphics[width=\textwidth]{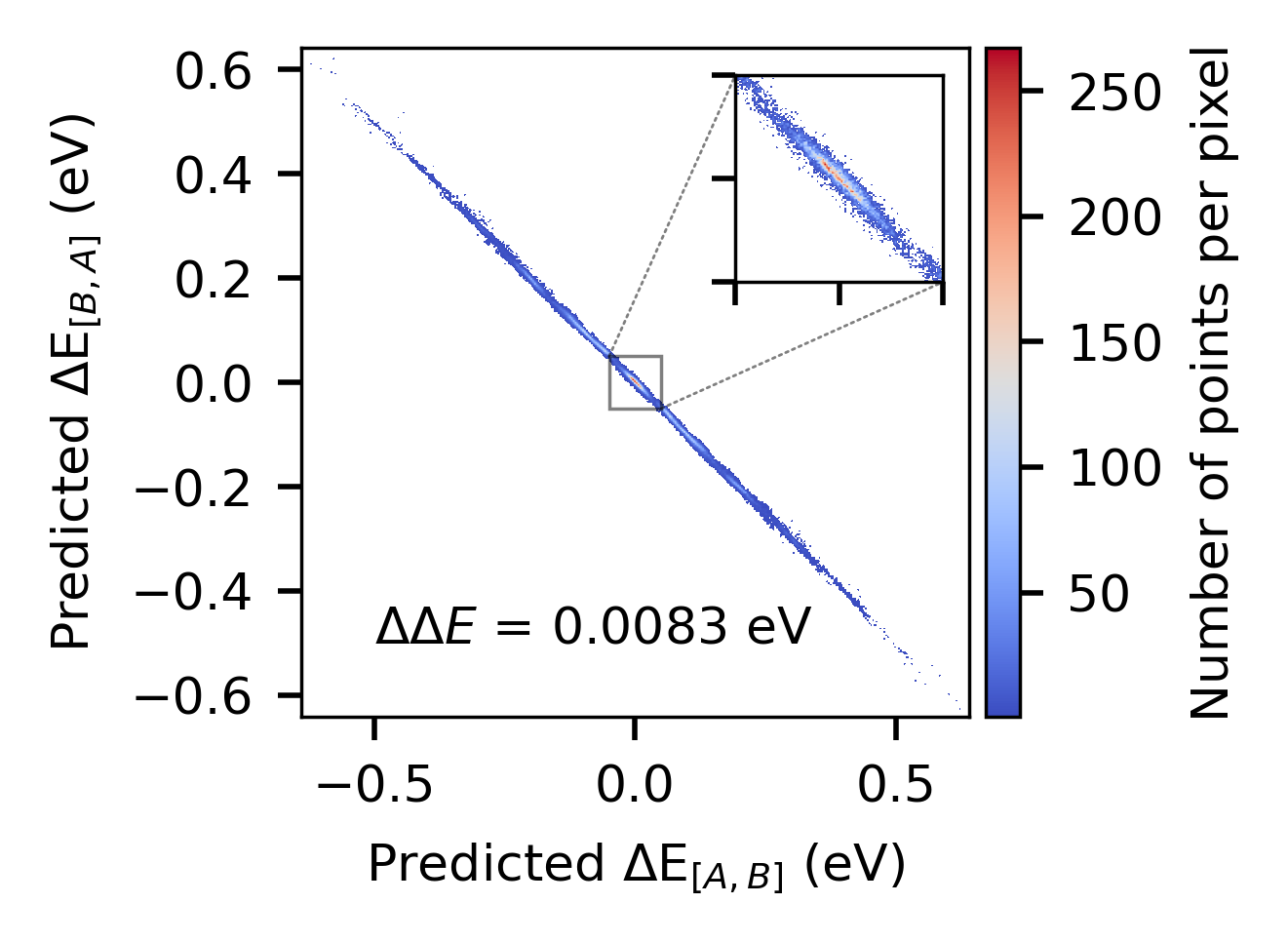}
      \caption{$\Delta \Delta_E$: Test}
    \end{subfigure}
    \caption{Performance of the \textit{soft anti-symmetry constraint} model $M_{SAS}$ to predict $\Delta E$. (a) $M_{SAS}$ predictions on 
    training data, b)  $M_{SAS}$ predictions on test data and c) Parity plot  comparing the $\Delta E$ predictions in the forward direction 
    ($\Delta E_{[A,B]}$) on the X-axis and the reverse direction ($\Delta E_{[B,A]}$) along the Y-axis. Inset shows the substantial deviation 
    from perfect detailed balance}
\label{fig:SC_results}
  \end{figure}
  
Training and test results for $M_{SAS}$ are indicated in Fig.~\ref{fig:SC_results}a \& b respectively. RMSE values of 0.0190 eV  and 0.0329 eV
are obtained on $D_{train}$ and $D_{test}$ respectively. Fig.~\ref{fig:SC_results}c shows the performance of the model in describing detailed 
balance. We see that there is substantial error in the comparison of the predicted $\Delta E$ in the forward and reverse directions. 
A  value of 0.0083 eV for $\Delta \Delta E$ is obtained on $D_{test}$.  

\subsubsection*{\textit{Hard anti-symmetry constraint} model:$M_{HAS}$}

Strict anti-symmetry can be imposed through a generalization of $M_{NC}$ that uses the basic feature of the architecture of $M_{KOL}$ but using 
\textit{transition-wise} fingerprints instead of \textit{Kolmogorov} fingerprints as input. In this case,  two sets of fingerprints are generated
by concatenating $\FTWA$ and $\FTWB$ in forward and reverse order as done in $M_{NC}$ and $M_{SAS}$ and each is fed to identical but separate DNN models $M_{E}$ similar to $M_{KOL}$. The outputs $E_{A}$  and $E_{B}$ of these 
two identical DNNs are then subtracted to yield $\Delta E_{[A,B]}$. This choice guarantees exact anti-symmetry, as exchange of 
$A$ and $B$ will by construction lead to a sign change of the output. As mentioned above, it is important to stress that anti-symmetry is not a sufficient condition for the DBC, which will be shown below.
The schematic for this model architecture is presented in 
Table~\ref{tab:detailed_balance}. Since the anti-symmetry is enforced in the architecture, this model is trained by minimizing $f_{min}^{NC}$.

\begin{figure}[h!]
    \begin{subfigure}[t]{0.33\textwidth}
      \includegraphics[width=\textwidth]{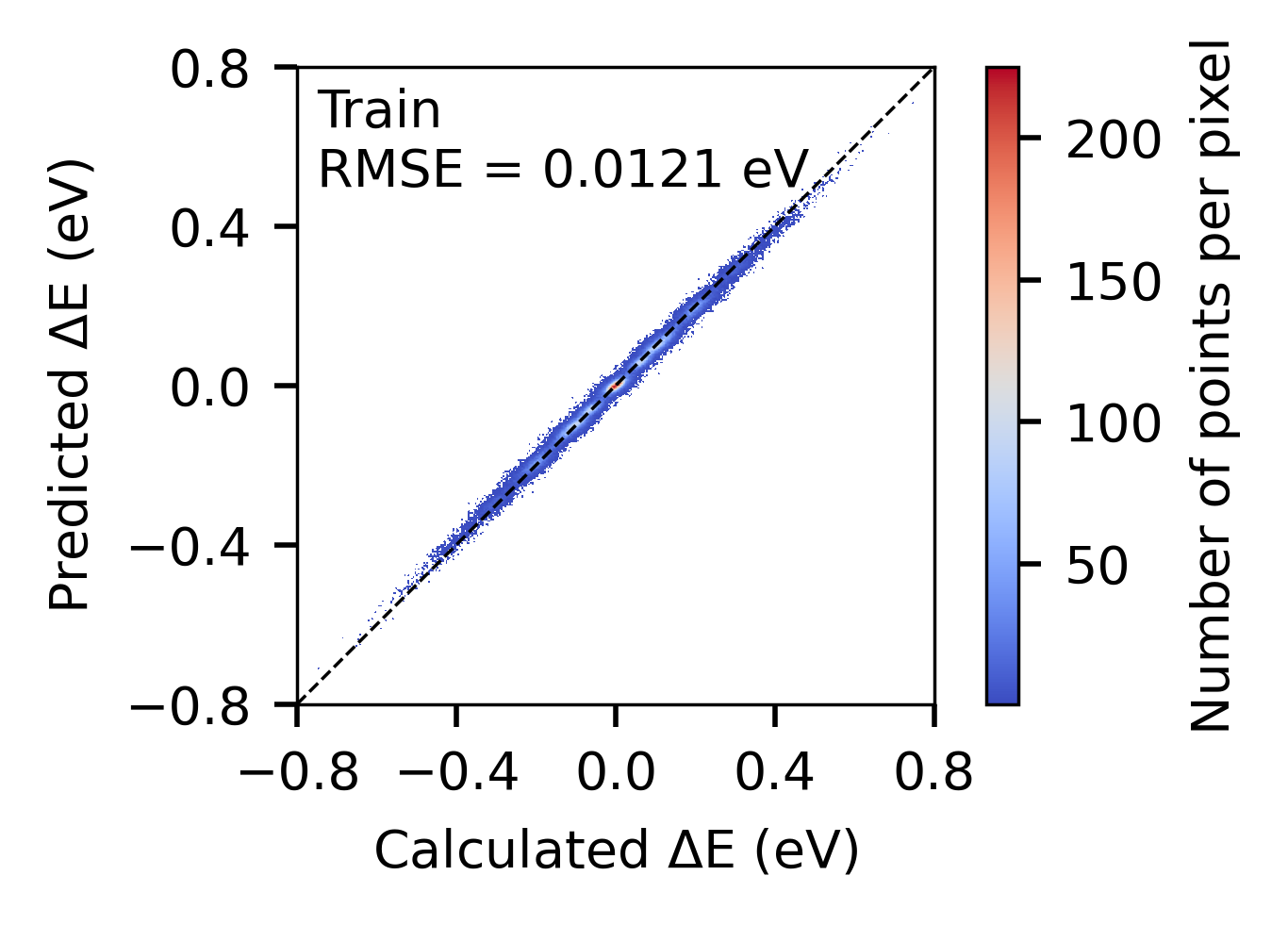}
      \caption{$\Delta_E$: Train}
    \end{subfigure}
    \hfill
    \begin{subfigure}[t]{0.33\textwidth}
      \includegraphics[width=\textwidth]{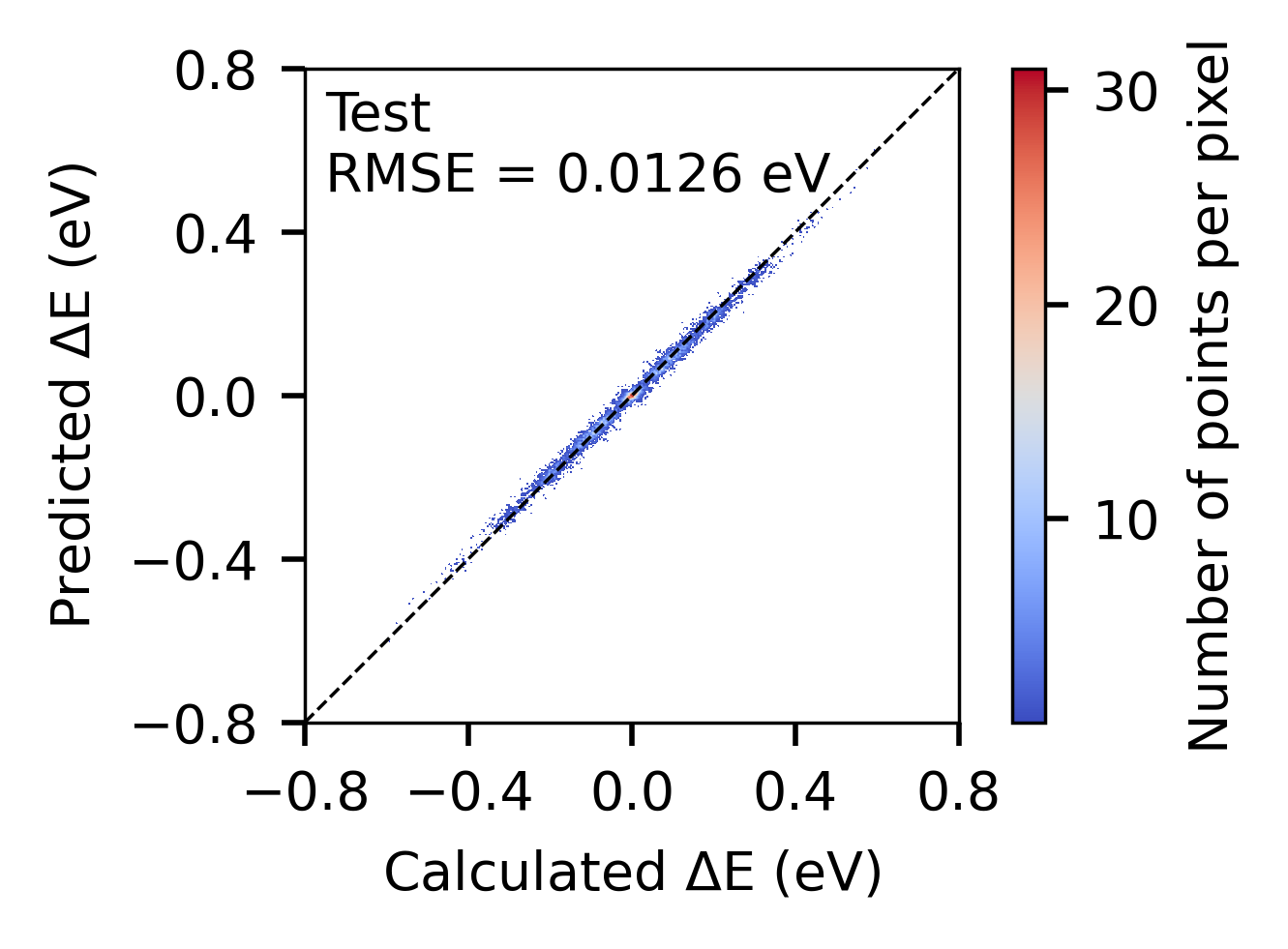}
      \caption{$\Delta_E$: Test}
    \end{subfigure}
    \begin{subfigure}[t]{0.33\textwidth}
      \includegraphics[width=\textwidth]{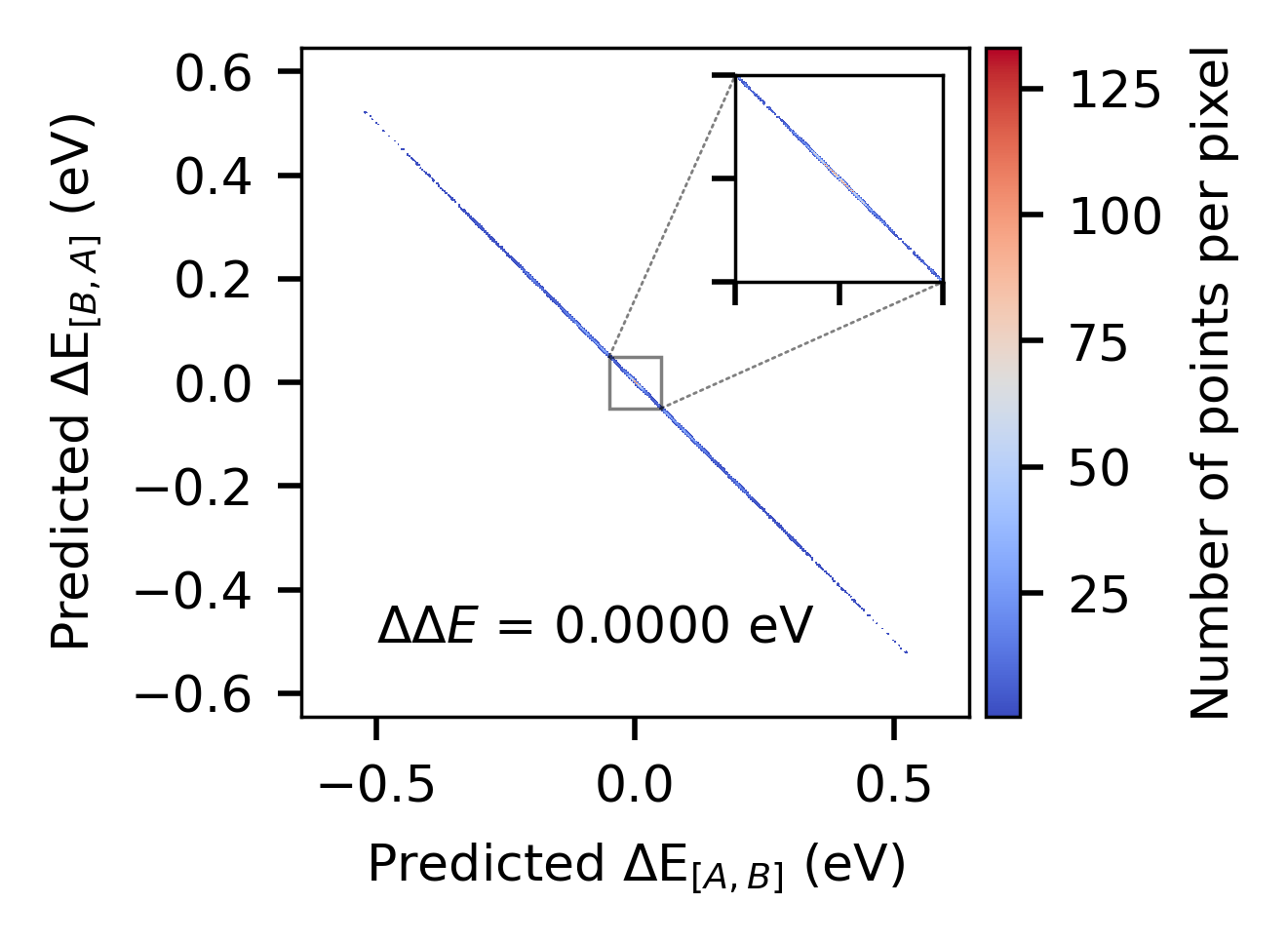}
      \caption{$\Delta \Delta_E$: Test}
    \end{subfigure}
    \caption{ Performance of the local \textit{hard anti-symmetry constraint} model $M_{HAS}$ to the prediction of $\Delta E$. (a) $M_{HAS}$ 
    predictions on training data, b)  $M_{HAS}$ predictions on test data and c) Parity plot comparing the $\Delta E$ predictions in the forward
     direction ($\Delta E_{[A,B]}$) on the X-axis and the reverse direction ($\Delta E_{[B,A]}$) along the Y-axis. Inset confirms that the 
     anti-symmetry of the energy difference ($\Delta \Delta E$ = 0.0000 eV) is exactly enforced by the architecture.}
\label{fig:LHC_results}
  \end{figure}
  
Training and test results for $M_{HAS}$ in predicting the magnitude of $\Delta E_{[A,B]}$ are reported in Fig.~\ref{fig:LHC_results}a \& b 
respectively. RMSE values of 0.0121 eV and 0.0126 eV are obtained on $D_{train}$ and $D_{test}$ respectively, a performance roughly on par 
with the soft anti-symmetry constraint model. However, $\Delta \Delta E$ is now identically zero by construction. 

These results as highlighted in Table~\ref{tab:perform_summ} clearly show that physics-informed ML architectures can overcome the trade-off inherent with multi-objective problems 
(here a combination of the accuracy in the prediction of the energy differences and of the extent to which the models obey physical constraints such as anti-symmetry or the DBC). Indeed, this tradeoff is apparent in the comparison of $M_{NC}$ and $M_{SAS}$, where $M_{SAS}$ shows a higher prediction error 
but a lower anti-symmetry error. In contrast, $M_{HAS}$, where exact anti-symmetry is enforced by the architecture, shows an even lower 
prediction error than both $M_{NC}$ and $M_{SAS}$. This trend further continues for $M_{KOL}$, where the Kolmogorov condition and the DBC 
are enforced by the architecture (an even stronger physical condition than local anti-symmetry), which demonstrates superior predictor 
performance than all other architectures. 

\begin{table}[h!]
\centering
\setlength\tabcolsep{20pt}
\renewcommand{\arraystretch}{1.5}
\begin{tabular}{cccc}
\hline
\multirow{2}{*}{Model} &
  \multicolumn{2}{c}{\begin{tabular}[c]{@{}c@{}}$\Delta E_{[A,B]} $\\ RMSE (eV)\end{tabular}} &
  \multirow{2}{*}{\begin{tabular}[c]{@{}c@{}}$\Delta \Delta E$ \\ (eV)\end{tabular}} \\ \cline{2-3}
  & \multicolumn{1}{c}{Train} & Test &   \hfill  \\ \hline
$M_{NC}$  & \multicolumn{1}{c}{0.0170}   & 0.0264   & 0.0127  \hfill \\ 
$M_{SAS}$ & \multicolumn{1}{c}{0.0190}   & 0.0329  &  0.0083  \hfill \\ 
$M_{HAS}$ & \multicolumn{1}{c}{0.0121}   & 0.0126  & 0.0000  \hfill \\ 
$M_{KOL}$ & \multicolumn{1}{c}{0.0091}   & 0.0098  & 0.0000   \hfill \\ \hline
\end{tabular}
\caption{Summary of results for the four DNN models to predict $\Delta E_{[A,B]}$ } and enforce the anti-symmetry condition $\Delta \Delta E = 0$.
\label{tab:perform_summ}
\end{table}

\section*{Discussion}

\subsection*{Effect of small training datasets}
\begin{figure}[h!]
    \begin{subfigure}[t]{0.45\textwidth}
      \includegraphics[width=\textwidth]{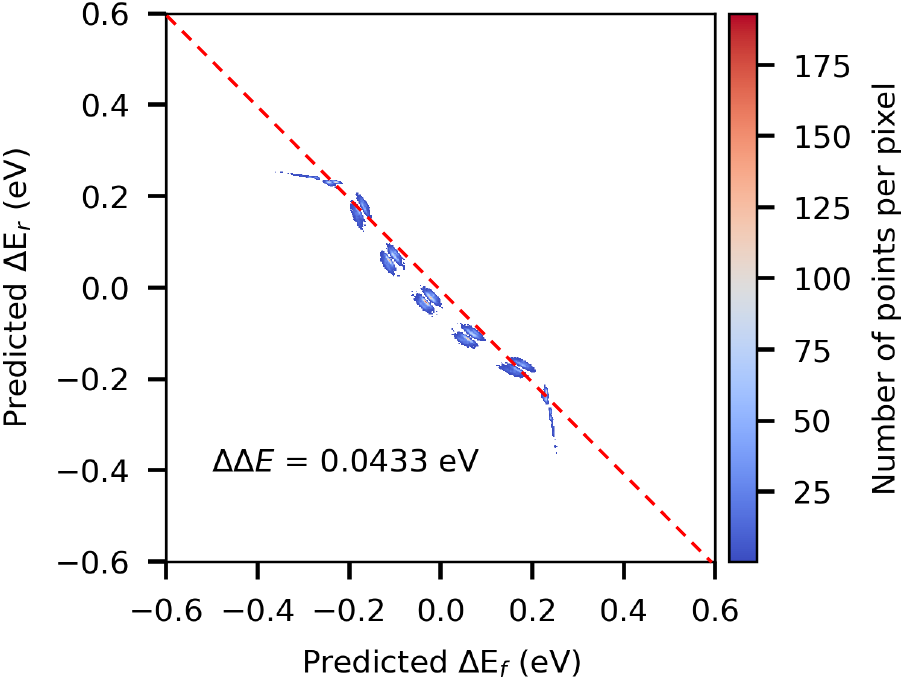}
      \caption{$M_{NC}$}
    \end{subfigure}
    \hfill
    \begin{subfigure}[t]{0.45\textwidth}
      \includegraphics[width=\textwidth]{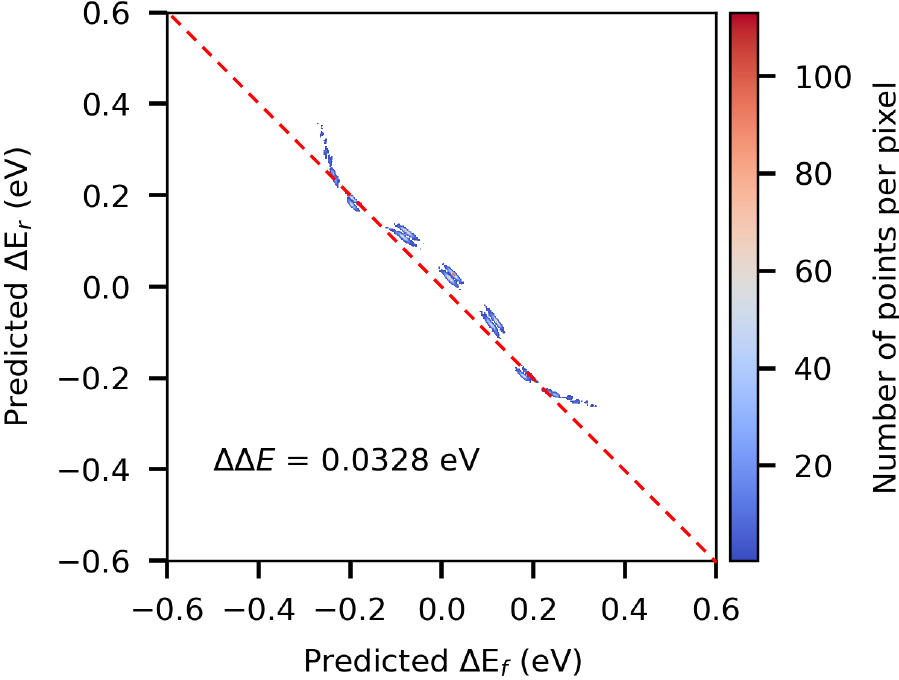}
      \caption{$M_{SAS}$}
    \end{subfigure}
    \begin{subfigure}[t]{0.45\textwidth}
      \includegraphics[width=\textwidth]{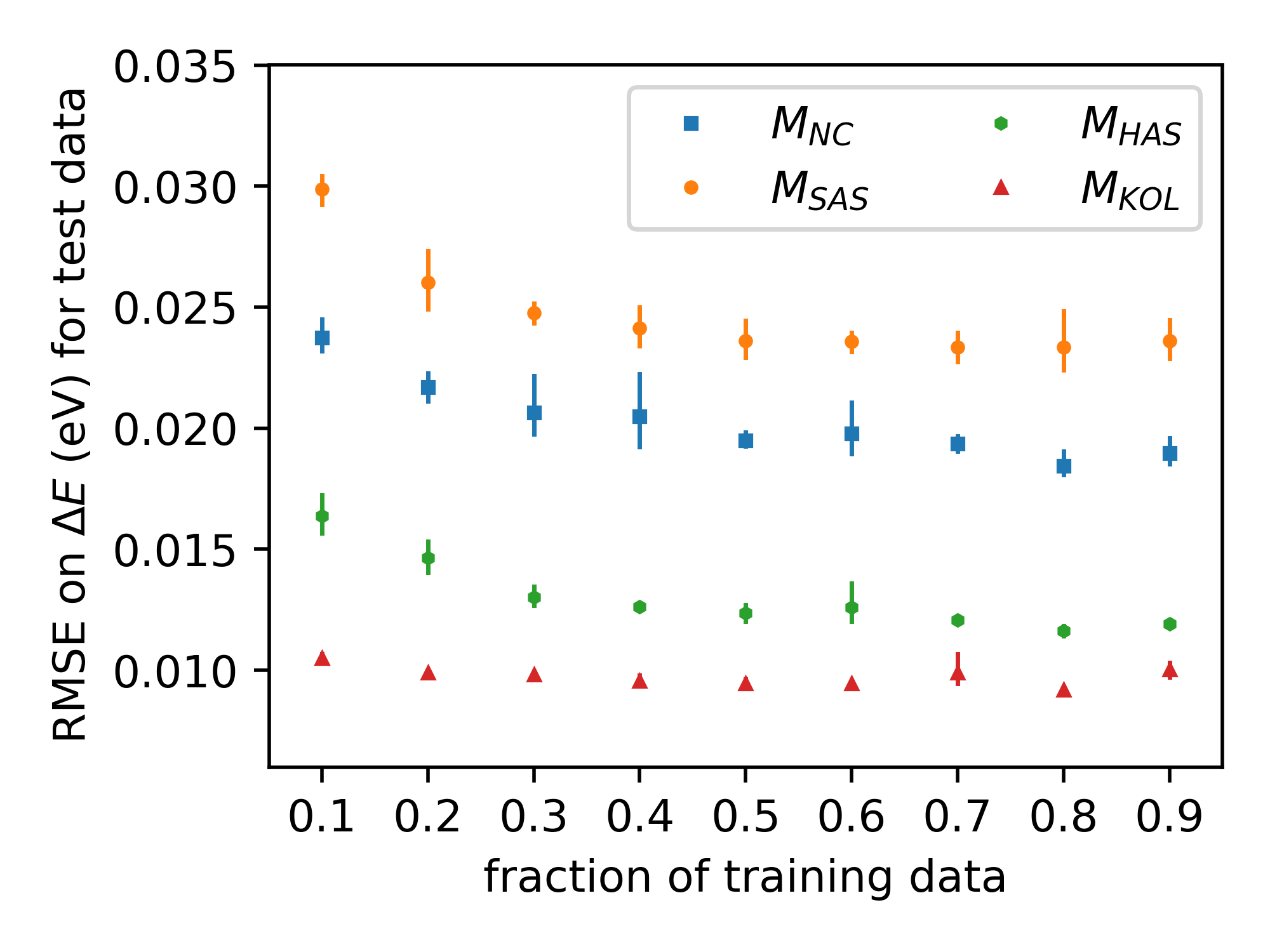}
      \caption{$\Delta E_{[A,B]}$ Vs. training data size}
    \end{subfigure}
        \hfill
    \begin{subfigure}[t]{0.45\textwidth}
      \includegraphics[width=\textwidth]{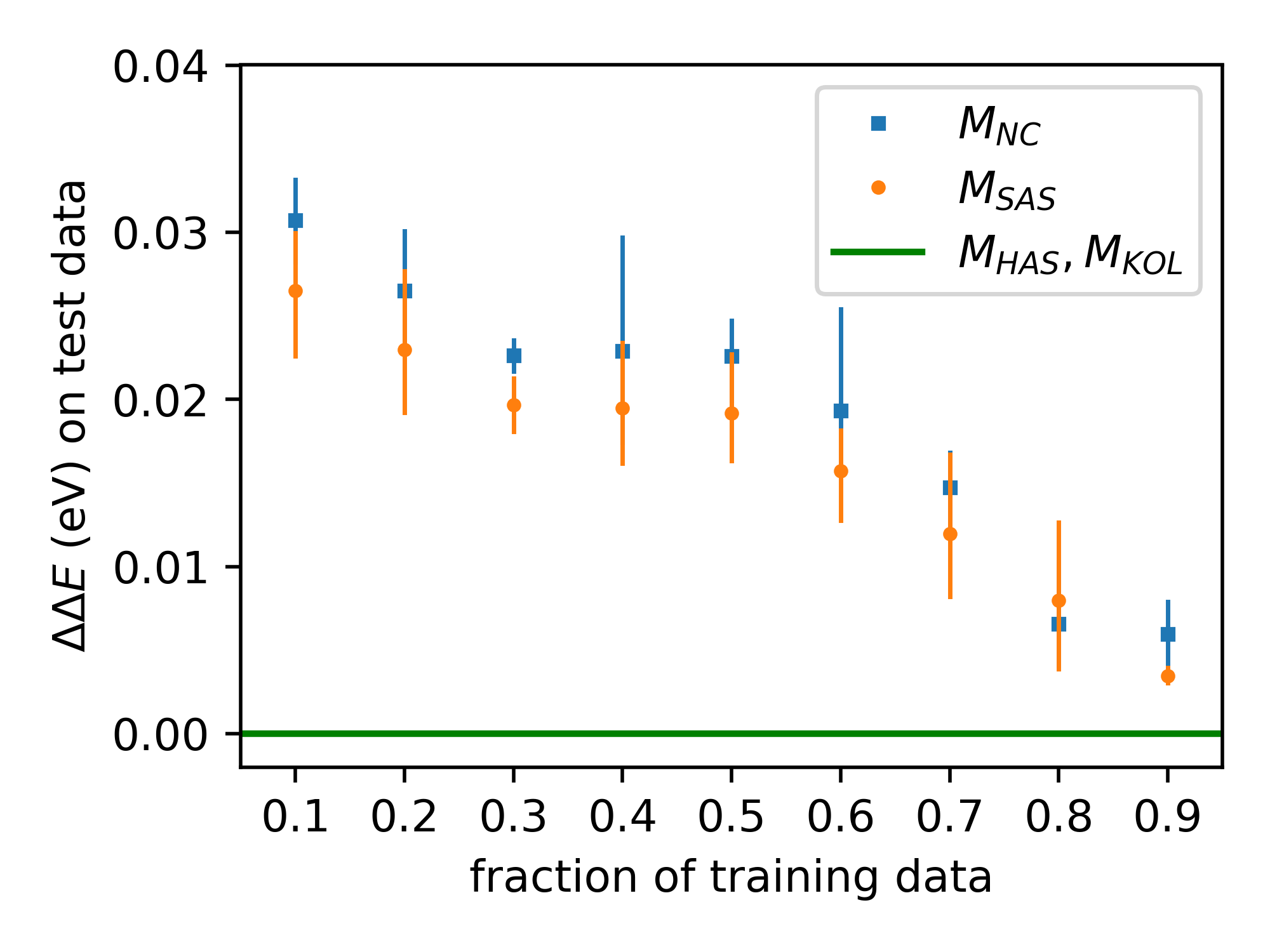}
      \caption{$\Delta \Delta E$ Vs. training data size}
    \end{subfigure}
    \caption{Effect of small training data size on $\delta E_{[A,B]}$ prediction and anti-symmetry. Parity plots for models trained on $D^{S}_{train}$ consisting of 1000 barriers  for (a) $M_{NC}$, b) $M_{SAS}$. c) Variation in $\delta E_{[A,B]}$ with increase in training data size, d)  in $\Delta E_{[A,B]}$ with increase in training data size, d) variation in $\Delta \Delta E_{[A,B]}$ with increase in training data size.}
    \label{fig:small_datasets}
  \end{figure}

In the $M_{HAS}$ and $M_{KOL}$ models for $\Delta E$, the anti-symmetry condition is encoded into the model architecture itself and thus its performance 
in terms of obeying detailed balance is expected to be independent of the size of the training dataset.
These two models thus guarantee perfect anti-symmetry even when trained on very small datasets. To demonstrate, we train the $M_{NC}$,
 $M_{SAS}$ models on a smaller dataset $D^{S}_{train}$ 
 containing only 1000 barriers (as compared to the original dataset which was trained on 
 $D_{train}$ consisting of around 22,500 barriers), the results of which are shown in Fig.~\ref{fig:small_datasets} a\&b. In Fig.~\ref{fig:small_datasets} a\& b, 
 we see that, when the training data is sparse, the detailed balance error is amplified for both the no-constraint and soft anti-symmetry constraint models. 
 In particular, for higher absolute values of $\Delta E$, the parity plots show increased and biased deviation from parity and these trends are opposite 
 for the no-constraint and soft anti-symmetry constraint cases. Critically, even in this regime of sparse data, the two models $M_{HAS}$ and $M_{KOL}$ 
 which have the anti-symmetry condition encoded into the model architecture itself still achieve perfect anti-symmetry. 

Fig.~\ref{fig:small_datasets} c\&d,  show the variations in  RMSE obtained for $\Delta E_{[A,B]}$ prediction and $\Delta \Delta E$ values achieved, respectively, 
for the four models with increasing training data size.  The RMSE values for $\Delta E{[A,B]}$ prediction saturate for all four models 
at around the 40\% training data mark. However, in Fig.~\ref{fig:small_datasets}d we see that, while for the $M_{HAS}$ and $M_{KOL}$ models the $\Delta \Delta E$ is identically 
zero for any size of training data, for the more traditional $M_{NC}$ and $M_{SAS}$ models, $\Delta \Delta E$ has yet to converge to $0$ for the dataset sizes considered here.

 \subsection*{Integration with ensemble closed-loop simulations}

 As mentioned earlier, anti-symmetry of the energy differences between states is not in itself sufficient to guarantee the DBC. Indeed, when translated to the task of predicting energy barriers and in conjunction with a definition of $\Delta E_{KRA}$ that is invariant to the exchange of initial and final states, the Kolmogorov criterion requires the stronger condition that the sum of the $\Delta E$ along {\em any} finite closed path be zero. 

\begin{figure}
    \begin{subfigure}[t]{0.45\textwidth}
      \includegraphics[width=\textwidth]{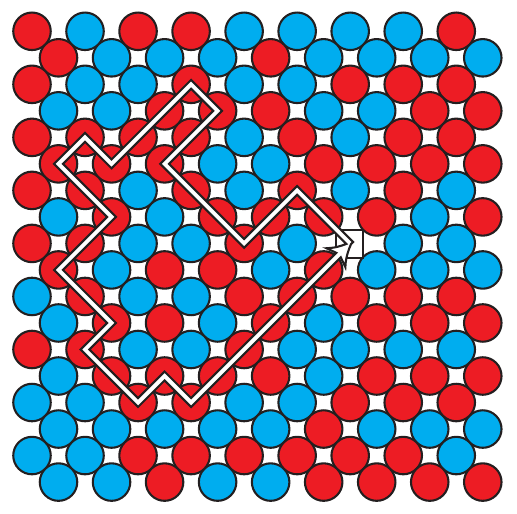}
      \caption{}
    \end{subfigure}
    \hfill
    \begin{subfigure}[t]{0.45\textwidth}
      \includegraphics[width=\textwidth]{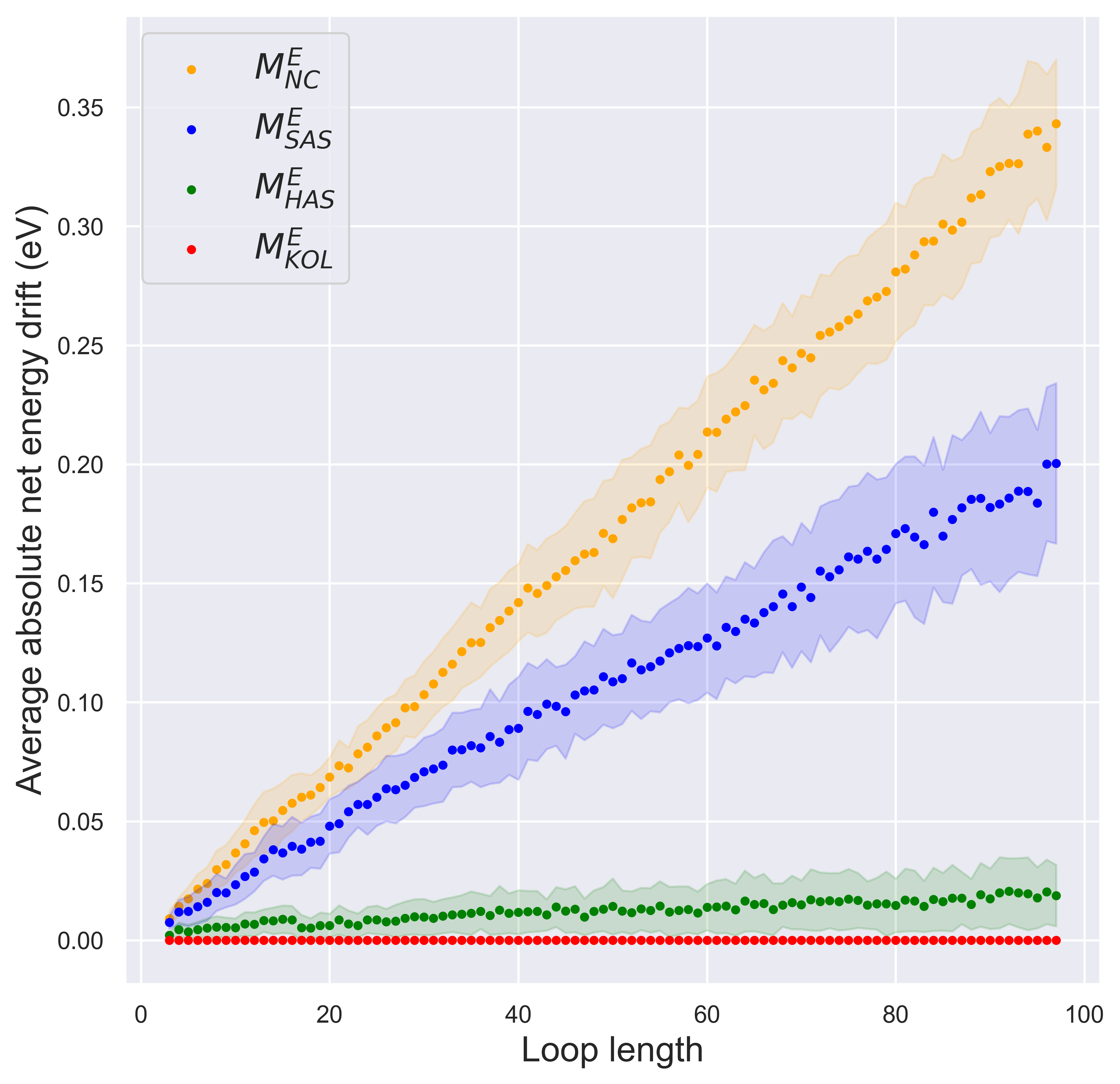}
      \caption{}
    \end{subfigure}
    \caption{(a) Schematic of a closed loop where the chemistry around the constructed loop is constant while the environments vary to maintain a 50:50 composition of Cu:Ni. (b) Average net energy loss computed for closed loop trajectories of varying length $L$ for the ensemble models of the four detailed balance implementations for $\Delta E$ prediction.}
\label{fig:KMC_integration}
  \end{figure}
  
  To demonstrate the importance of obeying the Kolmogorov criterion in these ML models, the four DNN models of $\Delta E_{[A,B]}$ for a vacancy hop in 
  the Cu-Ni binary alloy are evaluated with a calculation that quantifies the 
energy drift around closed loops in state-space, where a Kolmogorov-obeying model will show a zero net drift. 
For more robust predictions, we build ensemble models from multiple DNN instances for each of our four $\Delta E_{[A,B]}$ prediction model classes. 
 These provide a measure of uncertainty via a standard deviation value for every predicted value of $\Delta E_{[A,B]}$. We use the voting 
 ensemble method~\cite{dietterich2000ensemble} as implemented in the python package torchensemble.
 Voting trains $M$ base estimators independently, and the final prediction is taken as the average over the predictions from all base estimators. Here, 
 we use $M=10$ estimators to build our ensemble models denoted as $M^E$. Note that so averaging models that  obey the Kolmogorov criterion still yields an ensemble model that obeys the Kolmogorov criterion.

These results obtained are shown in Fig.~\ref{fig:KMC_integration}.
Each data point shows the average magnitude of the net energy $<|E_{net}(L)|>$, where the angled brackets denote an average calculated over
 100 independent randomly sampled geometric loops of length $L$. All lattice sites on a loop are occupied by a the same atomic species and a vacancy is cycled around
  every site of the loop so that the initial and final states are identical after a complete round-trip. Atomic species in the lattice that 
  do not lie along the loop are randomized - that is, the chemistry along the constructed loop is constant while the environment is randomly resampled. This is done 100 times at a 50:50 composition of Cu:Ni to create 100 independent loops in different random realizations of Cu:Ni. 

Random geometric loops are constructed through a heuristic sampling scheme which employs the NetworkX~\cite{hagberg2008exploring} python package.
 The scheme works by sampling from the cycle basis, or minimal collection of cycles (loops) that can ‘be ‘combined’ to form any possible loop 
 within a graph (crystal lattice). Loops are incrementally constructed by randomly choosing elements of the cycle basis and including them 
 if there is one overlapping edge with any part of the combined loop. The ‘combination’ is an XOR operation on the edges of the basis cycles 
 that are included. This process is repeated until a desired number of random loops have been generated for a specified range of lengths. The 
 average absolute net energy drift is then calculated by accumulating the energy differences predicted from each of the four models for each 
 nearest-neighbor lattice step the vacancy takes through the loop in one direction. This whole process is repeated for loops of varying sizes 
 from $L=3$ to $100$.

From Fig.~\ref{fig:KMC_integration}, it is apparent that $M^E_{KOL}$ models perform the best,
showing zero drift. The $M^E_{HAS}$ comes in second, with a small but noticeable drift that increases roughly linearly with loop size. This supports the prediction that single-step anti-symmetry is not sufficient to obey the DBC, as it does not guarantee that the Kolmogorov condition is obeyed along longer paths. This drift is even larger for the traditional no-constraint ($M^E_{NC}$) and soft anti-symmetry constraint ($M^E_{SAS}$) models, accumulating drifts of around 0.36 eV and 0.2 eV, respectively, for a vacancy loop consisting of 100 consecutive hops. 

These results show that violations of the Kolmogorov criterion can dramatically compound along long paths, even if anti-symmetry violations 
remain small for each individual transition. While it is {\em a priori} difficult to precisely quantify the potential deleterious effects of such 
violations on configurational properties inferred from the simulations, it appears equally difficult to {\em a priori} exclude that either 
serious or subtle pathologies could be introduced into long-time predictions. In absence of a compelling reason to do so (e.g., certain kinetics 
that violate detailed balance but obey global balance can preserve correct thermodynamics while accelerating relaxation to equilibrium 
\cite{bernard2009event}), it appears generally advisable to benefit from the strong guarantees that the DBC provides, especially when the accuracy of the predictions is also improved in the process.

\section*{Conclusions}

The prediction of transition barriers or of rate constants is a common and computationally expensive task in materials science and chemistry. As such, it has been recognized as an important target for machine learning approaches that can bypass the computationally intensive steps of direct methods. While physically-motivated constraints such as the detailed balance conditions can often be naturally enforced in traditional approaches, this has not been the case for ML approaches introduced thus far. In this paper, we show how the mathematical structure of rates obtained by transition state theory which guarantee that the DBC is obeyed can also be transferred to ML approaches by i) decomposing the overall rate prediction problem into two subtasks with well-defined invariance/anti-symmetry constraints with respect to the exchange of initial and final states, and ii) simple conditions on the featurization and architectures that can be employed for each task. This results in ML approaches that obey the DBC by construction, even in the low data limit. By comparing the performance of a method in this class to that of different variants that do not strictly obey the DBC, we show that the strict enforcement of physical constraints in fact does not entail a reduction in prediction accuracy, but, to the contrary, leads to lower prediction errors, in contrast to approaches that attempt at enforcing the anti-symmetry through the loss function, which entails a tradeoff between accuracy and anti-symmetry.
Therefore, this work highlights that well-designed physics-informed ML models can achieve the best of both worlds, simultaneously providing strong theoretical guarantees while also delivering better predictions. 

\section*{Methods}
\label{sec:methods}
\subsection*{Training Dataset Generation}
The training dataset of ${\Delta}E$ and kinetically resolved activation barriers ${\Delta}E^{KRA}$ was generated using the LAMMPS\cite{lammps} molecular dynamics code and an embedded-atom method (EAM)\cite{daw1984embedded} interatomic potentials for Cu-Ni alloys obtained following Ref.\ ~\cite{foiles1985calculation}.

 To generate a large database of ${\Delta}E_{[A,B]}$ and $E^{KRA}_{(A,B)}$ for a diverse set of local atomic configurations, a 4X4X4 face-centered cubic (FCC) supercell consisting of 255 atoms and a single vacancy was considered. A random neighbor pair, say ($\alpha$,$\beta$), is selected from a pristine supercell containing 256 atoms. Subsequently, a vacancy is introduced at lattice site $\beta$ by removing the atom, forming the initial structure. The atom initially occupying site $\alpha$ is relocated to site $\beta$, creating the final structure following the vacancy hop. To capture variations in atomic distributions around the vacancy sites, the remaining atomic sites are randomly populated with compositions of 50$\%$ Cu-50$\%$ Ni, 25$\%$ Cu-75$\%$ Ni, and 75$\%$ Cu-25$\%$ Ni. For each composition, $\approx$ 10,000 structures are generated. For each composition of the alloy, a lattice parameter corresponding to  Vegard's law between pure Cu and pure Ni is used and kept fixed during the respective simulations. Both initial and final points are then fully relaxed. The migration barrier is computed using 
  the climbing image nudged elastic band (CI-NEB) algorithm\cite{henkelman2000climbing}  with 5 intermediate images and a force convergence criterion of 1$\times$10$^{-2}$ eV/{\AA}. While relaxed structures are used in the calculation of all energies and barriers, the original unrelaxed structures were also recorded for use in training the ML models. We do this because lattice kMC models are evolved on idealized lattices.  In what follows, all the fingerprints will thus be generated using unrelaxed geometries. The models have been tested using fingerprints generated from relaxed geometries and while that improves the model performance, it complicates their translation to lattice kMC simulations.

 The ML models were trained on 75\%, validated on 15\%, and then tested on the remaining 10\% of the calculated data. The three Cu-Ni compositions considered were equally distributed among the training, validation and test datasets.

\subsection*{Environment Representation}
A prerequisite to learn any material property, including the target defect formation and migration energies using a machine learning algorithm, is the ability to encode the local configurational environments in a compact and expressive numerical fingerprint. Such fingerprints are generally chosen to obey key physical symmetries and invariances (such as rigid rotation and translation or permutation of like atoms) and to be continuous and differentiable with respect to atomic positions. Development of effective fingerprints for learning molecular and materials properties has been an active field of research for more than a decade now and a number of numerical representation schemes have been proposed, including symmetry functions,\cite{behlerGeneralizedNeuralNetworkRepresentation2007, behlerMetadynamicsSimulationsHighPressure2008, behlerRepresentingPotentialEnergy2014b} and closely related variants, such as those used in the  Accurate Neural networK engINe for Molecular Energies (ANAKIN-ME or ANI) framework \cite{s.smithANI1ExtensibleNeural2017, gaoTorchANIFreeOpen2020a}; bispectra of neighborhood atomic densities\cite{bartokGaussianApproximationPotentials2010b}; Coulomb matrices and related descriptors \cite{chmielaMachineLearningAccurate2017a, ruppFastAccurateModeling2012e}; smooth overlap of atomic positions (SOAP) \cite{bartokRepresentingChemicalEnvironments2013d,szlachtaAccuracyTransferabilityGaussian2014a, bartokGaussianApproximationPotentials2015};  graph based representations\cite{parkDevelopingImprovedCrystal2020, chenGraphNetworksUniversal2019, xieCrystalGraphConvolutional2018b, choudharyAtomisticLineGraph2021}; and others \cite{dussonAtomicClusterExpansion2022, jindalSphericalHarmonicsBased2017a,zongDevelopingInteratomicPotential2018b, botuMachineLearningForce2017b}. More details can be found in several recent reviews  \cite{ramprasadMachineLearningMaterials2017m, butlerMachineLearningMolecular2018b, pilaniaMachineLearningMaterials2021b, choudharyRecentAdvancesApplications2022}.

In this work, we adopt the fingerprinting scheme developed within the ANI framework to encode the atomic environment vector (AEV) corresponding to a local chemical environment around a reference atom or a point defect site. Each AEV is composed of radial and angular parts. The radial $AEV$ is further divided into sub-AEVs according to the atomic constituents forming the material. Likewise, the angular AEV is  composed of sub-AEVs for all unique triplets of elemental species. More specifically, for an $N$-component material, the AVE is formed by concatenating $N$ radial sub-AEVs and $\frac{1}{2}N(N + 1)$ angular sub-AEVs. To encode the local radial environment for an atom $i$ for element-type $X$, the components of a radial sub-AEV $G_m^R$ are constructed as:

\begin{equation}
\label{radial-AEV}
G_m^R (X) = \sum_{j \neq i}^{\substack{All~atom\\species~X}} \exp\left[- \eta(R_{ij}-R_s)^2 \right]  f_C(R_{ij}).
\end{equation}

\noindent Here the sum runs over all atoms of the elemental species $X$ and $R_{ij}$ represents the Euclidean distance between the central atom (or an arbitrary reference point) $i$ and an atom $j$. The index $m$ can, in principle, run over a two dimensional fine grid covering a set of discrete values for the hyperparameters $\eta$ and $R_s$. However, in an ANI fingerprint, only a single value of $\eta$ is chosen in conjunctions with  multiple $R_s$, producing thin Gaussian peaks that probe the local radial chemical environments outward from the atomic center. The parameter $\eta$ is used to tune the width of the Gaussian distribution while the $R_s$ allows for shifting of the center of the peak. The fingerprints are made spatially local by introducing a cutoff function 
\begin{equation}
\label{cutoff-function}
f_C(R_{ij}) =
    \begin{cases}
      \frac{1}{2} \cos(\frac{\pi R_{ij}}{R_C}) + \frac{1}{2}  &  \forall
 R_{ij}\leqslant R_C \\
     0 
    \end{cases}
\end{equation}

\noindent that ensures that the AEVs smoothly decay to zero beyond a cutoff distance  $R_C$. Radial sub-AEV are then generated by combining a set of radial symmetry functions $G_m^R (X)$, evaluated over a set M =\{$m_1$, $m_2$, $m_3$,...\} = \{($\eta$, $R_{s1}$), ($\eta$, $R_{s2}$), ($\eta$, $R_{s3}$),...\}.

The angular sub-AEV component $G_m'^A (X,Y)$ for an atom pair belonging to elemental species X and/or Y is similarly devised to capture the local angular environment around a central atom i with indices j and k running over neighboring atom pairs.

\begin{equation}
\label{angular-AEV}
G_m'^A (X,Y) =\sum_{j,k \neq i}^{\mathclap{\substack{All~atom\\ pairs~species\\X~Y}}} \frac { \left( 1 + cos (\theta_{ijk} - \theta_{s} ) \right)^{\zeta} }{ 2^{1-\zeta} }   \exp\left[ {-\eta\left(\frac{R_{ij}+R_{ik}}{2}-R_s\right)^2} \right] f_C(R_{ij}) f_C(R_{ik})
\end{equation}

\noindent The Gaussian term combined with the two cutoff functions again allows for exploitation of spatial locality. The index $m'$ runs over four separate parameters, namely, $\zeta$ $\theta_s$, $\eta$, and $R_s$. The latter two serve a similar purpose as discussed above in the context of Eq. \ref{radial-AEV}. The $\zeta$ parameter controls width of the peaks in the angular environment and variation in the $\theta_s$ parameter allows for probing of specific regions of the angular environment, much like in the case of $R_s$ for the radial part. In the angular sub-AEV, components are devised by sampling over multiple pairs of ($\theta_s$, $R_s$) on a two dimensional grid, while $\zeta$ and $\eta$ are kept fixed. The final fingerprint vector $\FTWA$ for a local state $A$ is given by a concatenation of all the radial and angular sub-AEVs.

\begin{figure}[h!]
\centering
\includegraphics[height=2.2in]{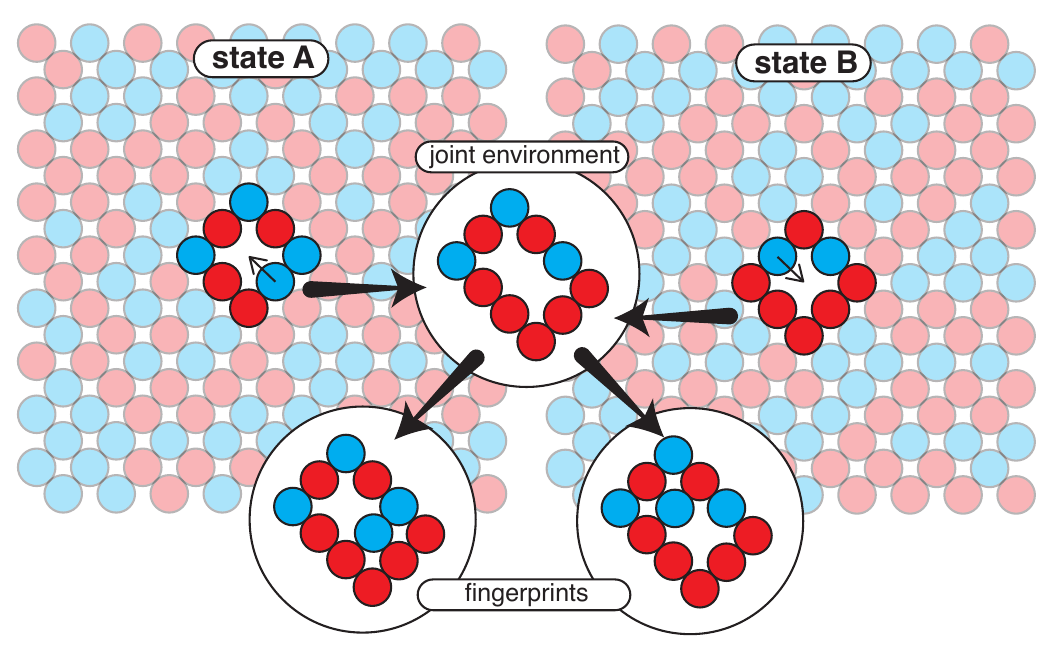}
\caption {Atomic environment selection for  vacancy migration. For the $E^{KRA}_{(A,B)}$ model and the no-constraint $M_{NC}$, soft anti-symmetry constraint $M_{SAS}$, and hard anti-symmetry constraint $M_{HAS}$ implementations of the $\Delta E_{[A,B]}$ models, 
the initial and final atomic configurations within a designated cutoff radius were combined to create an effective environment centered at 
the saddle point of the vacancy migration path (\textit{transition-wise}). For the 
\textit{Kolmogorov} implementation, the individual atomic environments centered at each individual vacancy site were chosen to extract 
fingerprints}
\label{fig:aev_config}
\end{figure}

Two types of fingerprints are considered. 
{\em Kolmogorov} fingerprints are obtained by first isolating a local environment around the target vacancy by selecting all atoms up to a cutoff 
radius $r_{cut}$ from the initial unrelaxed position of the vacancy
and computing the AEV completely independently for each state. In this case, the fingerprints of the initial state $A$ are completely agnostic to possible final states $B$ and vice-versa. In contrast, {\em transition-wise} fingerprints are constructed by first forming the union of the local environments in both initial and final states of the vacancy before computing the AEVs for each state. As shown in Fig~\ref{fig:aev_config} this leads to fingerprints that, e.g., contain information about the final location of the vacancy in the featurization of the initial state (and vice-versa).

\begin{figure}
  \centering
  \begin{minipage}[t]{.59\linewidth}
    \subcaptionbox{}
      {\includegraphics[width=\linewidth]{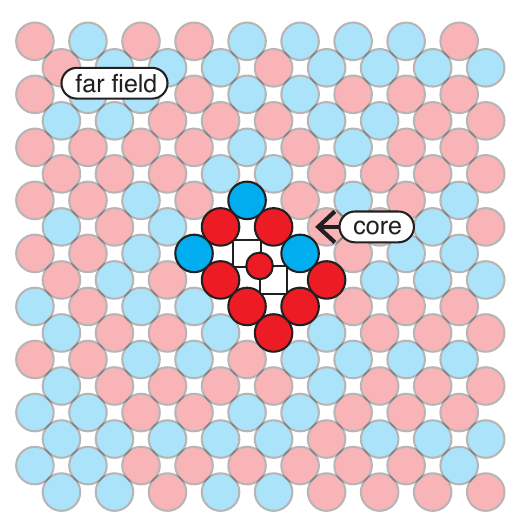}}%
  \end{minipage}%
  \hfill
  \begin{minipage}[b]{.39\linewidth}
    \subcaptionbox{}
      {\includegraphics[width=.9\linewidth]{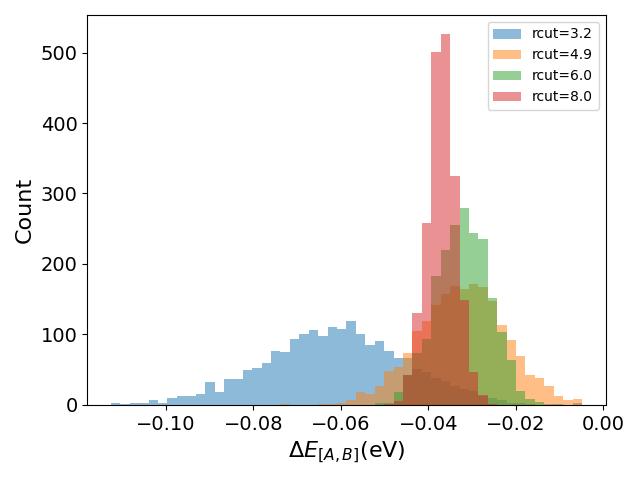}}
    \subcaptionbox{}
      {\includegraphics[width=.9\linewidth]{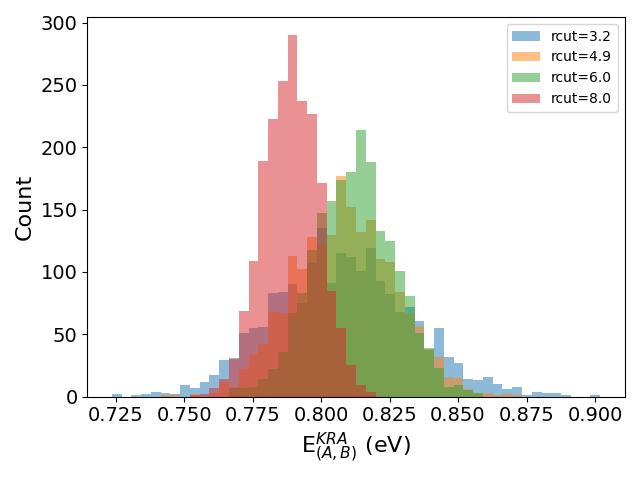}}%
  \end{minipage}
  \caption{(a) Schematic of core and far-field for a binary alloy and Migration energy statistics obtained by fixing core atoms upto different values of $r_{cut}$ for $Cu_{50}-Ni_{50}$.  in a  $6X6X6$ supercell decoupled into (b) $\Delta E_{[A,B]}$ and (c) $\Delta E^{KRA}_{(A,B)}$.}
  \label{fig:rcut_variation}
\end{figure}

The choice of $r_{cut}$ results from a tradeoff between computational cost and accuracy. This tradeoff was explored by explicitly computing energy barriers in different environments that differ only by their composition {\em outside} of the transition-wise environments defined by a given value of $r_{cut}$. The composition of this "far field" region was then randomly resampled 2000 times for each cutoff radius. Fig.~\ref{fig:rcut_variation}a reports both the distribution of exact energy differences $\Delta E_{[A,B]}$ and of kinetically-resolved barriers $\Delta E^{KRA}_{(A,B)}$ for different values of $r_{cut}$. The width of the distribution of $\Delta E_{[A,B]}$, representing an intrinsic "noise" level that no ML model can resolve using features computed at the given $r_{cut}$, decreases from $\approx$0.1 eV to 0.02 eV as the cutoff is increased from 3.2 \r(A) to 8.0 \r(A). This indicates a significant proportion of the errors observed above, which correspond to models trained with $r_{cut}$ = 6 \r{A}, can be attributed to the influence of the unresolved environment.

An analysis of the distribution of $\Delta E^{KRA}_{(A,B)}$ shows a slightly larger effect of the environment, with an observed width of about 0.025 eV at 8.0 \r(A). For $r_{cut}$ = 6 \r{A}, the width of the distribution was about 0.035 eV, again similar to the error observed above. These results indicate that increasing the value of $r_{cut}$ likely offers a simple path toward initially improving accuracy, although architectural details will eventually dominate the error. 

\subsection*{Model architecture}
Deep neural network (DNN) models were built to learn and predict the relationship between the local configurational environment and i) the energy difference between two local minima of the vacancy  ($\Delta E$) and ii) the kinetically resolved activation barrier ($\Delta E_{KRA}$)
for the migration. All these models are trained by the Adam algorithm~\cite{kingma2014adam} implemented in PyTorch within a python framework. Linear layers with the rectified linear unit (ReLU)~\cite{agarap2018deep} activation function were employed. 
The number of hidden layers was optimized using the training and validation data and drop-out layers and early-stopping were employed to avoid overfitting. 

\section*{Data Availability}
Data related to this work is available on reasonable request.
\section*{Code Availability}
The source code used in this study is available upon request. 

\section*{Acknowledgements}
The authors thank Ju Li (MIT) for asking a question that motivated the development of the Kolmogorov model. Research presented in this paper was supported by the Laboratory Directed Research and Development program of Los Alamos National Laboratory under project number 20220063DR. Computational support for this work was provided by LANLs high-performance computing clusters. This work was supported by the U.S. Department of Energy through the Los Alamos National Laboratory. Los Alamos National Laboratory is operated by Triad National Security, LLC, for the National Nuclear Security Administration of U.S. Department of Energy (Contract No. 89233218CNA000001).

\section*{Author contributions statement}
Blas Uberuaga, Danny Perez and Ghanshyam Pilania proposed and supervised the entire project. Anjana Talapatra worked on the development, testing of the Machine Learning models. Danny Perez carried out the mathematical analysis of the ML architectures. Blas Uberuaga, Danny Perez, Ghanshyam Pilania and Anjana Talapatra  analyzed and discussed results. Anup Pandey performed all the LAMMPS simulations to generate the training data for the Machine Learning models. Matthew Wilson and Ying Wai Li developed the ensemble closed loop framework. Anjana Talapatra prepared the final draft of the manuscript which was then reviewed and edited by all authors.

\section*{Competing Interests}
The authors declare that there are no competing interests.


  \end{document}